\documentclass[a4paper,fleqn,usenatbib]{mnras}
\usepackage[dvipsnames,usenames]{color}
\usepackage{colortbl}
\usepackage{mathptmx}
\usepackage{graphicx}
\usepackage{amsmath}
\usepackage{natbib}
\newcommand{\etal}{{et al}\/.}
\newcommand\rah{\mbox{$^{\mathrm h}$}}%
\newcommand\ram{\mbox{$^{\mathrm m}$}}%
\begin{document}
\pubyear{2015}
\title[Pictor A]{Deep {\it Chandra} observations of Pictor A}
\author[M.J.\ Hardcastle \etal]
{M.J.\ Hardcastle$^1$\thanks{E-mail: m.j.hardcastle@herts.ac.uk}, E.\ Lenc$^{2,3}$, M.\ Birkinshaw$^{4}$, J.H.\ Croston$^{5,6}$,
J.L. Goodger$^1$\thanks{Present address: Information Services, City
  University London, Northampton Square, London EC1V 0HB, UK},
\newauthor H.L.\ Marshall$^7$, E.S.\ Perlman$^8$,
A.\ Siemiginowska$^9$, \L. Stawarz$^{10}$ and D.M. Worrall$^{4}$\\
$^1$ School of Physics, Astronomy and Mathematics, University of
  Hertfordshire, College Lane, Hatfield AL10 9AB, UK\\
$^{2}$ Sydney Institute for Astronomy, School of Physics, The University of Sydney, NSW 2006, Australia\\
$^{3}$ ARC Centre of Excellence for All-Sky Astrophysics (CAASTRO), Redfern, NSW 2016, Australia\\
$^4$ H.H.\ Wills Physics Laboratory, University of Bristol, Tyndall Avenue,
  Bristol BS8 1TL, UK\\
$^5$ School of Physics and Astronomy, University of Southampton,
  Southampton SO17 1BJ, UK\\
$^6$ Institute of Continuing Education, University of Cambridge, Madingley
  Hall, Madingley, CB23 8AQ, UK\\
$^7$ Kavli Institute for Astrophysics and Space Research, Massachusetts
Institute of Technology, 77 Massachusetts Ave., Cambridge, MA 02139,
USA\\
$^8$ Department of Physics and Space Sciences, Florida Institute of Technology, 150 W. University Blvd., Melbourne, FL 32901, USA\\
$^9$ Harvard-Smithsonian Center for Astrophysics, 60 Garden Street,
Cambridge, MA 02138, USA\\
$^{10}$ Astronomical Observatory, Jagiellonian University, ul. Orla 171, 30-244 Krak\'ow, Poland\\
}
\maketitle
\begin{abstract}
We report on deep {\it Chandra} observations of the nearby broad-line
radio galaxy Pictor A, which we combine with new Australia Telescope
Compact Array (ATCA) observations. The new X-ray data have a factor 4
more exposure than observations previously presented and span a
15-year time baseline, allowing a detailed study of the spatial,
temporal and spectral properties of the AGN, jet, hotspot and lobes.
We present evidence for further time variation of the jet, though the
flare that we reported in previous work remains the most significantly
detected time-varying feature. We also confirm previous tentative
evidence for a faint counterjet. Based on the radio through X-ray
spectrum of the jet and its detailed spatial structure, and on the
properties of the counterjet, we argue that inverse-Compton models can
be conclusively rejected, and propose that the X-ray emission from the
jet is synchrotron emission from particles accelerated in the boundary
layer of a relativistic jet. For the first time, we find evidence that
the bright western hotspot is also time-varying in X-rays, and we
connect this to the small-scale structure in the hotspot seen in
high-resolution radio observations. The new data allow us to confirm
that the spectrum of the lobes is in good agreement with the
predictions of an inverse-Compton model and we show that the data
favour models in which the filaments seen in the radio images are
predominantly the result of spatial variation of magnetic fields in
the presence of a relatively uniform electron distribution.
\end{abstract}

\section{Introduction}
\label{sec:intro}

\subsection{X-ray jets}

One of the most important and unexpected discoveries of {\it Chandra}
has been the detection of X-ray emission from the jets of a wide range
of different types of radio-loud AGN (see
\citealt{Harris+Krawczynski06} and \citealt{Worrall09} for reviews).
In \cite{Fanaroff+Riley74} class I (FRI) radio galaxies, including
nearby objects like Cen A (e.g. \citealt{Hardcastle+07-4}) or M87
\citep[e.g.][]{Harris+03} the jet X-ray emission is believed to be due
to the synchrotron mechanism. In this case the X-rays trace electrons
with TeV energies and radiative loss lifetimes of years, and so can
give us crucial insights into the location and the nature of particle
acceleration in these sources. Dynamical modelling of FRI jets
suggests that the particle acceleration regions are associated with
bulk deceleration as the jets slow from relativistic to mildly
relativistic speeds \citep[e.g.][]{Laing+Bridle02}. It is possible, in
some jets, that a significant fraction of the particle acceleration
giving rise to X-ray emission is the result of shocks in the jet
related to its interaction with the stellar winds of its internal
stars \citep{Wykes+15}. In more powerful FRI jets, like that of M87,
this process is probably not energetically capable of producing all
the X-ray emission, and instead internal shocks due to jet variability
\citep[e.g.,][]{Rees78b} or jet instabilities driving shocks and
turbulence \citep{Bicknell+Begelman96,Nakamura+Meier14}, still
associated with bulk deceleration, may be required.

Our understanding of the X-ray emission from the jets of more powerful
radio AGN, including `classical double' FRII radio galaxies and
quasars, is much more limited. A wide variety of X-ray counterparts to
jets have been seen, ranging from weak X-ray emission from localized
`jet knots' in radio galaxies like 3C\,403 \citep{Kraft+05} or 3C\,353
\citep{Kataoka+08} to bright, continuous structures extending over
hundreds of kpc in projection, as seen in the prototype of the class,
PKS 0637$-$752 \citep{Schwartz+00}. Two mechanisms have been invoked
to explain the X-ray emission from powerful jets. The first is
inverse-Compton scattering of the CMB (hereafter IC/CMB) by a
population of low-energy electrons \citep{Tavecchio+00,Celotti+01}.
This model relies on high bulk Lorentz factors $\Gamma \ga 10$ and
small angles to the line of sight in order to produce detectable
X-rays; it has been applied successfully to the bright, continuous
X-ray jets in many core-dominated quasars, but has a number of
problems in explaining all the observations, particularly the
broad-band SED and the spatial variation of the radio/X-ray ratio
\citep{Hardcastle06}, the observed knotty jet X-ray morphology
\citep{Tavecchio+03,Stawarz+04}, the non-detection of the gamma rays
predicted in the model
\citep{Georganopoulos+06,Meyer+Georganopoulos14,Meyer+15}, and the
high degree of observed optical/ultra-violet polarization
\citep{Cara+13}. The second process is synchrotron emission, which
does not depend on large jet Doppler factors but does require {\it in
  situ} particle acceleration as in the FRIs. This model is more often
applied to weak X-ray `knots' seen in jets (and counterjets) of radio
galaxies, and at present has the weakness that it cannot explain {\it
  why} there is localized particle acceleration at certain points in
the jets, since, unlike the case of the FRIs, there appears to be no
preferred location for X-ray emission, and certainly no association
with jet deceleration. (Indeed, there is no direct evidence for
  significant jet bulk deceleration in FRII jets at all, with the
  exception of the possible and debatable evidence provided by the
  X-rays themselves \citep{Hardcastle06}, and on theoretical grounds
  the interpretation of the hotspots as jet termination shocks implies
  supersonic bulk jet motion with respect to the internal jet sound
  speed.) Detailed studies of individual objects are required to
determine how and where the two X-ray emission processes are
operating.

The X-ray jet of the FRII radio galaxy Pictor A (\citealt*{Wilson+01},
hereafter W01) provides a vital link between the two extreme classes
of source discussed above. Like those of the powerful core-dominated
quasars, Pic A's jet extends for over 100 kpc in projection, and is
visible all the way from the core to the terminal hotspot. However, as
the source is a lobe-dominated broad-line radio galaxy, its brighter
jet is expected to be aligned towards us ($\theta \la 45^\circ$) but
{\it not} to be within a few degrees of the line of sight; {\it a
  priori} we would not expect significant IC/CMB X-rays. (A small jet
angle to the line of sight would imply a very large, Mpc or larger,
physical size for the source.) In addition, the existing {\it Chandra}
data show that the bright region of the jet has a steep spectrum
(\citealt{Hardcastle+Croston05}, hereafter HC05) and there is a faint
but clear X-ray counterjet, neither of which would be expected in
IC/CMB models. If the X-rays in Pic A are indeed synchrotron in
origin, then it provides us with an opportunity to investigate how a
powerful FRII source can accelerate particles along the entire length
of its jet. Pic A is also a key object because of its proximity; at
$z=0.035$ it is one of the closest FRIIs, and the closest
example of a continuous, 100-kpc-scale X-ray jet. Thus we can
investigate the fine structure in the jet, key to tests of {\it all}
possible models of the X-ray emission, at a level not possible in any
other powerful object.

Pic A was observed twice in the early part of the {\it Chandra}
mission. A 26-ks observation taken in 2000 provided the first
detection of the X-ray jet (W01). In 2002 a 96-ks observation of the
X-ray bright W hotspot was taken: these data were used by HC05 in
their study of the lobes (see below). In 2009, we re-analysed these
data in preparation for a study of the jet and found clear evidence at
around the $3\sigma$ level for {\it variability} in discrete regions
of the jet between these two epochs: we obtained a new observation
which strengthened the evidence for variability in the brightest
feature, 34 (projected) kpc from the core, to the $3.4\sigma$ level
after accounting for trials (\citealt{Marshall+10}, hereafter M10).
Another feature at 49 kpc from the core was found to be variable at
the $\sim 3\sigma$ level. The discovery of X-ray variability in the jet of
Pic A, the first time it had been seen in an FRII jet, was a
remarkable and completely unexpected result which has very significant
implications for our understanding of particle acceleration in FRII
jets in general. It requires that a significant component of the X-ray
emission (and thus the particle acceleration, in a synchrotron model)
comes from very small, pc-scale, features embedded in the broader jet.
Variability is in principle expected in synchrotron models of X-ray
jets, since the synchrotron loss timescales are often very short,
implying short lifetimes for discrete features in `impulsive' particle
acceleration models. However, the nearby X-ray synchrotron jets in the
FRIs Cen A and M87 have been extensively monitored, and most features
show little or no evidence for {\it strong} variability
\citep[e.g.][]{Goodger+10}, suggesting that particle acceleration in
these jets is generally long-lasting on timescales much longer than
the loss timescale. A dramatic exception is the HST-1 knot in the
inner jet of M87, which {\it Chandra} has observed to increase in
brightness by a factor $\sim 50$ on a timescale of years
\citep{Harris+06,Harris+09}. HST-1 in M87 may provide the closest known analogue
of what we appear to be seeing in Pic A, but the flares in Pic A are
both much more luminous and much further from the AGN. Again, there is
no reason to suppose that Pic A is unique among FRII radio galaxies,
but, as the closest and brightest of FRII X-ray jets, it provides our
best chance of understanding the phenomenon, and it may also provide
insight into the presumably related variability on kpc spatial scales
that is starting to be seen in gamma rays from lensed blazars
\citep{Barnacka+15}.

\subsection{Hotspots and lobes}
\label{sec:intro-lobes}

Pic A's proximity, radio power, and lack of a rich environment
emitting thermal X-rays make it a uniquely interesting target in
X-rays in several other ways. With the possible exception of Cygnus A
\citep{Hardcastle+Croston10}, where thermal emission from the host
cluster is dominant and inverse-Compton emission is hard to detect
reliably in the X-ray, it is the brightest lobe inverse-Compton source
in the sky: for FRIIs lobe inverse-Compton flux scales roughly with
low-frequency radio flux, so this is a direct result of its status as
the second brightest FRII radio galaxy in the sky at low frequencies
\citep{Robertson73}. Because of this, the inverse-Compton lobes have
been extensively studied in earlier work
(W01; \citealt{Grandi+03}; HC05; \citealt{Migliori+07}). It also
hosts the brightest X-ray hotspot known \citep[e.g.,
  W01;][]{Hardcastle+04,Tingay+08}. Thus a deep {\it Chandra}
observation of the whole source allows us to study the spatially
resolved X-ray spectrum of the lobes and hotspot to a depth not
possible in any other FRII. Key questions here are what the spectra of
the lobes and hotspot actually are -- relatively few sources even
provide enough counts to estimate a photon index -- and how well they
agree with the predictions from the inverse-Compton and synchrotron
models for the lobe and hotspot respectively. In addition, in the case
of the lobes, we can use spatially resolved images of the
inverse-Compton flux to study the (projected) variation of magnetic
field and electron number density in the lobes, as discussed by HC05
and \cite{Migliori+07}.

\subsection{This paper}

In this paper we report on the results of a {\it Chandra} multi-cycle
observing programme, carried out since the results reported by
M10, targeting the inner jet of Pic A. As we shall see in more detail
below, this gives a combined exposure on the source of 464 ks, nearly
a factor 4 improvement in exposure time with respect to the last
large-scale study of the source by HC05 (though the sensitivity is not
improved by such a large factor, as the sensitivity of the ACIS-S
continues to drop with time), and a factor 16 improvement in exposure
time since the original analysis of the jet by W01. In addition, the
new data give us a long time baseline, sampling a range of different
timescales and comprising 9 epochs spread over 15 years, with which to
search for temporal variability in the jet and other components of the
source. We use this new dataset to investigate the spatial, temporal
and spectral properties of the X-ray emission from all components of
the radio galaxy.

We take the redshift of Pic A to be 0.0350 and assume $H_0 = 70$ km
s$^{-1}$, $\Omega_{\rm m} = 0.3$ and $\Omega_\Lambda = 0.7$. This
gives a luminosity distance to the source of 154 Mpc and an angular
scale of 0.697 kpc arcsec$^{-1}$. Spectral fits all take into account
a Galactic column density assumed to be $4.12 \times 10^{20}$
cm$^{-2}$. The spectral index $\alpha$ is defined in the sense $S_\nu
\propto \nu^{-\alpha}$, where $S_\nu$ is the flux density, and so the
photon index $\Gamma = 1 + \alpha$. Errors quoted are $1\sigma$ (68
per cent confidence) statistical errors unless otherwise stated (see
discussion of calibration errors in Section \ref{sec:analysis}).

\section{Observations and data processing}

\subsection{X-ray}

\begin{table}
\caption{Details of the {\it Chandra} observations of Pictor A. The 13
observations used in the paper are listed together with their
observation date, duration, pointing position, satellite roll angle
and epoch number (observations with the same epoch number are combined
when variability is considered).}
\label{tab:obs}
\begin{tabular}{llllrr}
\hline
Obs. ID&Date&Exposure&Pointing&Satellite&Epoch\\
&&(ks)&&roll (deg)\\
\hline
346 & 2000-01-18 & 25.8 & Core&322.4&1\\
3090 & 2002-09-17 & 46.4 & W hotspot&88.1&2 \\
4369 & 2002-09-22 & 49.1 & W hotspot&88.1&2\\
12039 & 2009-12-07 & 23.7 & Jet &3.2&3\\
12040 & 2009-12-09 & 17.3 & Jet &3.2&3\\
11586 & 2009-12-12 & 14.3 & Jet &3.2&3 \\
14357 & 2012-06-17 & 49.3 & Jet &174.3&4\\
14221 & 2012-11-06 & 37.5 & Jet &36.2&5\\
15580 & 2012-11-08 & 10.5 & Jet &36.2&5\\
15593 & 2013-08-23 & 49.3 & Jet &110.5&6\\
14222 & 2014-01-17 & 45.4 & Jet &322.6&7\\
14223 & 2014-04-21 & 50.1 & Jet &232.7&8\\
16478 & 2015-01-09 & 26.8 & Jet &315.2&9\\
17574 & 2015-01-10 & 18.6 & Jet &315.2&9\\
\hline
\end{tabular}
\end{table}

As discussed in Section \ref{sec:intro}, {\it Chandra} has observed Pic A
for a useful duration\footnote{We do not make use of two very short
  exposures taken early in the mission.} on 14 separate occasions over
the past 14 years, for a total of 464 ks of observing time. Details of
the observations are given in Table \ref{tab:obs}.

The pointings of the observations differ, and this affects the quality
of the available data on various regions of the source. The original
observations (obsid 346) were pointed at the active nucleus, with a
roll angle which included both lobes on the ACIS-S detector. The 2002
observations (3090 and 4369) were pointed at the W hotspot, and much
of the E lobe emission was off the detector, as discussed by HC05. All
our subsequent observations (2009-2014) have had the aim point about 1
arcmin along the jet in the W lobe, but roll angle constraints have
been applied so that the E lobe always lies on the S3 or S2 chips, and
also to avoid interaction of the readout streak from the bright
nucleus with any important features of the source. Because the W lobe
is generally on the S3 chip, which has higher sensitivity, and also
because of the missing 2002 data, the observations of the E lobe are
roughly 2/3 the sensitivity of those of the W lobe. However, the new
observations are still a great improvement in sensitivity terms on the
data available to HC05. The different pointing positions mean that the
effective point spread function of the combined dataset is a
complicated function of position, and we comment on this where it
affects the analysis later in the paper.

\begin{figure*}
\includegraphics[width=1.0\linewidth]{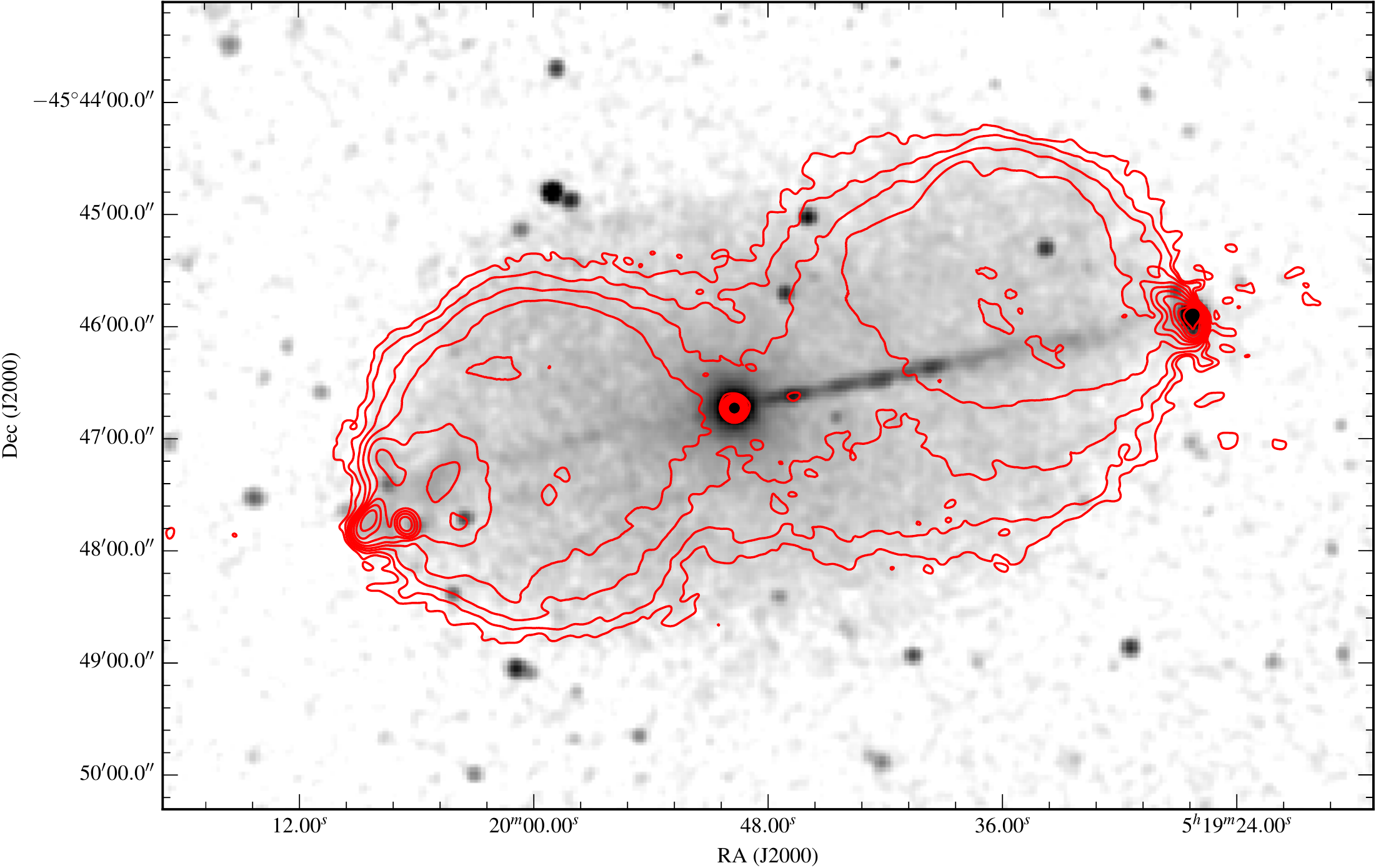}
\caption{X-ray emission from Pictor A and its field. The greyscale
  shows an exposure-corrected image made from all the data in the 0.5-5.0 keV
  passband, smoothed with a Gaussian with a FWHM of 4.6 arcsec and
  with a logarithmic transfer function to highlight fainter structures.
  Superposed are contours of our ATCA 5.5-GHz image tapered to a
  resolution of 5 arcsec: contour levels are at $0.6 \times
  (1,2,4\dots)$ mJy beam$^{-1}$. }
\label{fig:all}
\end{figure*}

The data were all reprocessed in the standard manner using {\sc ciao}
4.7 (using the {\it chandra-repro} script) and CALDB 4.6.7. The
readout streaks were removed for each obsevation and the events
files were then reprojected to a single physical co-ordinate system
(using observation 12040 as a reference). The {\it merge\_obs} script was used to produce merged events
files and also to generate exposure maps and exposure-corrected (`fluxed')
images, which are used in what follows when images of large regions of
the source are presented: images of raw counts in the merged images
are shown when we consider compact structure, for which local
variations in the instrument response can be neglected. Spectra were
extracted from the individual events files using the {\it specextract}
script, after masking out point sources detected with {\it
  celldetect}, and subsequently merged using the {\it
  combine\_spectra} script. Weighted responses were also generated
using {\it specextract}. Spectral fitting was done in {\sc xspec} and
{\sc sherpa}.

Fig.\ \ref{fig:all} shows an exposure-corrected image of the centre of the field
covered by the observations.

\subsection{ATCA observations}
\label{sec:obs-atca}

Pictor A was observed in 2009 with the Australia Telescope Compact
Array (ATCA) in three separate observations: two in 6-km
configurations and one in a compact 352-m configuration, as
summarized in Table \ref{table:atcaobs}. The Compact Array Broadband
Backend (CABB; \citet{Wilson+11}) was used with a correlator cycle
time of 10 s and the full 2048-MHz bandwidth (as 1-MHz channels) centred
at 5.5 GHz (6 cm) and 9.0 GHz (3 cm). The primary beam of the ATCA
varies from 9 arcmin to 13 arcmin full width at half-maximum (FWHM)
across the full 2 GHz band at 6 cm. The overall extent of Pictor A is
$\sim8$ arcmin and so the source is completely contained within the
primary beam over the entire 6-cm frequency range. Unfortunately the
3-cm band is not of practical use as bright components of Pictor A
fall outside the 6--7-arcmin FWHM of the primary beam in this band, so
our results here use the 5.5-GHz data only.

\begin{table}
\caption{Details of ATCA CABB observations of Pictor A.}
\label{table:atcaobs}
\begin{tabular}{llrr}
\hline
Obs. Date & ATCA config. & Central Frequency & Time on-source \\
 &  & (GHz, GHz) & (hr) \\
\hline
2009-06-16 & 6A    & 5.5, 9.0 & 12.0 \\
2009-08-29 & 6D    & 5.5, 9.0 & 13.5 \\
2009-12-06 & EW352 & 5.5, 9.0 & 10.5 \\ [1ex]
\hline
\end{tabular}
\end{table}

The three observations, when combined using multi-frequency synthesis (MFS),
provide near-complete $uv$-coverage spanning from 450 $\lambda$ out to
128 k$\lambda$. While the shortest baselines provide sensitivity to
structures as large as 7.5 arcmin in extent, the longest baselines
provide a resolution approaching 1.6 arcsec. In the work
presented here a $uv$ taper between 70 and 100 k$\lambda$ has been
applied, giving resolutions of 1.7-2.2 arcsec, to highlight the
large-scale structure of the source and to ensure that sufficient
visibility data (with overlapping baselines) is available to model the
spectral variation of these structures. A more heavily tapered image
is used to show details of the lobes (Fig.\ \ref{fig:all}).

To facilitate initial calibration, a 10 minute scan of the standard
flux density calibrator source PKS B1934-638 was made at each epoch
and a bright secondary calibrator, PKS B0537-441
($\alpha=5\rah38\ram50\fs36$, $\delta=-44^\circ 5' 8.94''$),
was observed for 4-6 minutes every 17-20 minutes during target
observations of Pictor A. While observing Pictor A, the pointing
centre was set to the position of the core from earlier ATCA
observations ($\alpha=5\rah19\ram49\fs75$,
$\delta=-45^\circ 46' 43.80''$) so that the hot spots and
lobes would be contained within the primary beam of the ATCA at 6 cm.

Primary flux density calibration, initial time dependent calibration
and data flagging for radio frequency interference was performed using
standard calibration procedures for the ATCA in the data reduction
package {\sc Miriad} \citep{Sault+95}. The three epochs were then combined,
frequency channels averaged to 8 MHz channels. Initial imaging in
{\sc Miriad} indicated that the deconvolution algorithms available in that
package were not adequate enough to deal with the complex spatial and
frequency-dependent structures present in Pictor A. The visibilities
were therefore exported into $uv$-FITS format so that they could be
processed with other packages.

Initial attempts to process the data with Common Astronomy Software
Applications (CASA\footnote{http://casa.nrao.edu/}), where more
advanced deconvolution algorithms were available, encountered issues
associated with the difficulty of simultaneously deconvolving and
calibrating data from an array with only a small number of baselines.
This prevented the standard clean-selfcal cycle from converging
towards a solution that accurately represented the source -
particularly around the hotspots where the synthesised-beam
side-lobes were difficult to separate from the diffuse lobe emission.

As an alternative approach, an attempt was made to perform
$uv$-visibility modelling of the different structures in the source.
To achieve this, the data were imported into {\sc Difmap}
\citep{Shepherd+94}. {\sc Difmap} allows components to be added and
modelled in the visibility domain while also modelling for simple
power-law spectral effects. The main structures of Pictor A (lobes,
hotspots and lobes) were iteratively modelled (for position, size,
orientation and spectral index) using a combination of Gaussian and
point-like components. As more source flux was recovered in the
component model, successive iterations of phase self-calibration were
performed between additional iterations of component modelling.
Ultimately, once a significant proportion of the source was recovered
in the component model, amplitude self-calibration was also performed.
At this stage it was clear that the calibration solutions at the hot
spots were different to those at the AGN core. The cause of this is
most likely due to pointing errors, which would shift the position of
the hotspot closer and further from the FWHM of the primary beam as a
function of time, and also the rotation of the primary beam over the
course of an observation (the ATCA has an Alt-Az mount and so the sky
rotates with respect to the feed over time). When peeling the
north-west hotspot, we see a smooth increase in scatter in the gain
corrections from around 1.5 per cent at the lower end of the band to
5.5 per cent at the top end of the band. While this correction
encapsulates both pointing errors and primary beam rotation errors (as
well as, in principle, any intrinsic variability in the source) it is
reasonably consistent with what one would expect with a pointing error
of $\sim 5$ arcsec rms, based on modelling the primary beam within
{\sc miriad}: this is significantly better than the worst-case
pointing errors of $\sim 15$ arcsec observed at the ATCA.

To minimize the observed position-dependent gain errors the technique
known as `peeling' \citep{Intema+09} was used to generate
position-dependent calibration solutions at the north-west hotspot and
at the AGN core. The technique involves iteratively subtracting the
sky model for everything except the direction of interest and then
determining the calibration solution for that direction. One of us
(EL) developed two software tools needed to do this for the ATCA data:
one to subtract {\sc difmap} components from a {\sc difmap} visibility
FITS file and another to compare an uncalibrated and calibrated {\sc
  difmap} FITS file to determine the gain corrections applied and then
transfers these to another {\sc difmap} FITS file. Applying the
peeling techniques improved the dynamic range of the resulting image
by more than a factor of 4 and resulted in a residual off-source image
noise of $\sim 40$ $\mu$Jy beam$^{-1}$. The final calibrated, modelled
and restored image, which is equivalent to the zero-order term of the
MFS imaging at a reference frequency of 5.5 GHz, was imported back
into {\sc Miriad} and the task \emph{linmos} used to correct for
primary beam attenuation. There is no reliable single-dish flux
measurement at 5.5 GHz, but we would expect a total source flux
density of 17.9 Jy based on interpolation between the Parkes catalogue
flux at 2.7 GHz and the 23-GHz WMAP data \citep{Bennett+13}, or 21.8
Jy if the 1.41-GHz Parkes data is used as the low-frequency point; our
final image contains 18.8 Jy, which sits well between these limits.
The low-resolution 5-GHz images of \cite{Perley+97} contain 23 Jy at
4.9 GHz, implying a flux difference of order 10 per cent after
correction for spectral index, but given flux calibration uncertainty
and the fact that both images are significantly affected by the
relevant telescope's primary beam, we do not regard this as
problematic.

The residual noise level of the final image is approximately a factor
of 5 higher than the estimated thermal noise for this observation but
still provides the highest dynamic range (46000:1) yet achieved for
this complex source. Further gains could potentially be obtained with
improved modelling and peeling of the north-west hot spot, as the
highest residual errors are still concentrated on this region. Such
improvements, however, are not required for the present analysis.

We have verified that the ATCA radio core, with a position in the new
images of $05\rah19\ram49\fs724$, $-45^\circ 46' 43.86''$, is aligned
with the peak of the {\it Chandra} emission from the active nucleus to
a precision of better than 0.1 arcsec; accordingly, we have not
altered the default astrometry of the {\it Chandra} and ATCA images.
There may well be some small astrometric offsets in the {\it Chandra}
data far from the aim point, which would effectively blur or smear the
{\it Chandra} PSF on these scales, but we see no evidence that they
are large enough in magnitude to affect our observations. Our radio
core position is in good agreement with the VLBI position of
$05\rah19\ram49\fs7229$, $-45^\circ 46' 43.853''$ quoted by
\cite{Petrov+11}.

\subsection{Other data}

The radio data of \cite{Perley+97} were kindly made available to us by
Rick Perley. The high-frequency, high-resolution VLA images do not
show the radio core (we have only images of the two hotspots,
sub-images of a larger image which is no longer available) and so we
cannot align them with the X-ray data manually, but the hotspot images
do not show any very large discrepancy with the ATCA data on visual
inspection. \cite{Perley+97} quote a core position from the short
baselines of their BnA-configuration
X-band data which in J2000.0
co-ordinates is $05\rah19\ram49\fs693$, $-45^\circ 46' 43.42''$,
0.55 arcsec away from our best position: however, this difference does
not necessarily affect all the data in the same way, and in any case the
images we use are probably also shifted as a result of phase self-calibration.

{\it Hubble Space Telescope} ({\it HST}) observations of the jet were
taken as part of this project, but are not described here: see Gentry
et al. (2015) for details of the observations and their results. We
comment on the implications of these observations for models of the
jet in the discussion.

\section{Analysis}
\label{sec:analysis}

\begin{table}
\caption{Approximate 0.5--5.0-keV counts (summed over all
  observations) in the key features of the radio galaxy seen in
  X-rays.}
\label{tab:counts}
\begin{tabular}{lrrll}
\hline
Component&Net counts&Error&Region used&Background\\
\hline
AGN & 119277 & 345 & Circle & Concentric \\
Jet & 7077 & 124 & Box & Adjacent \\
Counterjet & 490 & 61 & Box & Adjacent \\
W hotspot & 32464 & 182 & Circle & Concentric \\
E hotspot & 2092 & 105 & Ellipse & Concentric \\
Lobes & 40537 & 790 & Ellipse & Concentric \\
\hline
\end{tabular}
\end{table}

As Fig.\ \ref{fig:all} shows, many features of the radio galaxy are
detected in X-ray emission. In addition to the bright nucleus, we see
emission from the well-known jet and hotspot on the W side of the
source, with the jet now seen to extend all the way to the hotspot at
250 arcsec (174 kpc in projection) from the nucleus. A jet in the E
lobe (hereafter the `counterjet') is now clearly detected, although
much fainter than the jet, and appears to extend all the way from the
nucleus to an extended region of emission associated with the E
hotspots. Finally, the lobes of the radio galaxy are very clearly
detected, presenting very uniform surface brightness in X-rays with
some X-ray emission extending beyond the lowest radio contours. Table
\ref{tab:counts} lists the total number of {\it
  Chandra} counts in the combined observations in each of these
features in order to give an indication of the significance at which
they are detected and the degree of certainty with which we can
discuss their properties.

It should be noted that the number of counts obtained for some
components of the system ($>10^4$) puts us in the regime, discussed by
\cite{Drake+06}, in which calibration uncertainties are likely to
dominate over statistical ones. Methods to include calibration
uncertainties in the analysis of Chandra data have been discussed by,
e.g., \cite{Lee+11} and \cite{Xu+14}. As the uncertainties are rarely
critical to our analysis, particularly given that we construct time
series for all the brightest components of the source, we do not
make use of such methods, but we comment below qualitatively whenever
the calibration uncertainty is likely to exceed the quoted statistical
uncertainty. For a power-law fit, the typical calibration
uncertainties on the photon index are $\sigma_{\rm sys} \approx 0.04$
\citep{Drake+11}.

In the following subsections we discuss the properties and origins of
each of the X-ray features, comparing with data at other wavelengths
where appropriate. Table \ref{tab:spectra} gives a summary of the
properties of X-ray spectral fits for discrete regions of the
combined X-ray dataset.

\begin{table*}
\caption{X-ray spectra of discrete regions: spectral parameters and
  fitting statistic. Parameters given without errors are fixed in the
  fit. Symbols are as follows: $\Gamma_1$, photon index of the fitted
  power law; $\Gamma_2$, photon index of the high-energy part of a
  broken power-law model; $kT$, temperature of a thermal model; $E_{\rm G}$, rest energy of a Gaussian.}
\label{tab:spectra}
\begin{tabular}{llrrrrr}
\hline
Region&Model&Photon index ($\Gamma_1$)&Second parameter&PL 1-keV flux&$\chi^2$&d. o. f.\\
&&&(photon index or energy/keV)&density (nJy)&\\
\hline
AGN (annulus)&PL&$1.88 \pm 0.01$&&$1750 \pm 20$&454.3&326\\
AGN (annulus)&PL + Gaussian&$1.90 \pm 0.01$&$E_{\rm G}=6.36 \pm 0.02$ keV&$1760 \pm 20$&417.4&324\\[5pt]
Jet (inner)&PL&$1.92 \pm 0.03$&&$11.7 \pm 0.2$&134.5&144\\
Jet (outer)&PL&$1.96 \pm 0.09$&&$2.9 \pm 0.2$&39.1&48\\[5pt]
Counterjet&PL&$1.7 \pm 0.3$&&$1.7 \pm 0.3$&14.3&20\\[5pt]
W hotspot (entire)&PL&$1.94 \pm 0.01$&&$90.5 \pm 0.5$&394.1&304\\
&broken PL&$1.86 \pm 0.02$&$\Gamma_2 = 2.16_{-0.04}^{+0.06}$&$90.8 \pm 0.6$&329.4&302\\
&pure thermal&--&$kT = 3.14 \pm 0.05$&--&567.8&303\\
&PL + thermal&$2.01 \pm 0.05$&$kT = 4.0 \pm 0.4$ keV&$63 \pm 5$&346.3&302\\
W hotspot (compact)&PL&$1.97 \pm 0.01$&&$76.8 \pm 0.5$&418.7&283\\
&broken PL&$1.87 \pm 0.02$&$\Gamma_2 = 2.23_{-0.04}^{+0.07}$&$77.2 \pm 0.5$&341.1&281\\
W hotspot (bar)&PL&$1.83 \pm 0.03$&&$11.5 \pm 0.2$&142.4&144\\[5pt]
E hotspot (whole)&PL&$1.76 \pm 0.10$&&$7.4 \pm 0.4$&87.5&84\\
E hotspot (X1, X3 excluded)&PL&$1.80 \pm 0.12$&&$5.9 \pm
0.4$&55.3&65\\[5pt]
Lobe (whole)&PL&$1.57 \pm 0.04$&&$99 \pm 1$&127.4&108\\
Lobe (E end)&PL&$1.64 \pm 0.19$&&$7.5 \pm 0.9$&56.4&54\\
Lobe (E middle)&PL&$1.34 \pm 0.12$&&$30 \pm 1$&98.1&94\\
Lobe (middle)&PL&$1.67 \pm 0.08$&&$25 \pm 2$&85.4&104\\
Lobe (W middle)&PL&$1.75 \pm 0.08$&&$30 \pm 2$&112.2&105\\
Lobe (W end)&PL&$1.54 \pm 0.10$&&$16 \pm 1$&84.3&96\\
Lobe (outside contours)&PL&$2.07 \pm 0.15$&&$8.7 \pm 0.4$&54.6&69\\
&Thermal&--&$kT = 2.7 \pm 0.5$ keV&--&61.7&69\\
&PL + thermal&1.57&$kT = 0.33 \pm 0.07$ keV&$7.1 \pm 0.5$&56.5&68\\
\hline
\end{tabular}
\end{table*}
\subsection{The AGN}
\label{sec:AGN}

\begin{figure}
\includegraphics[width=1.0\linewidth]{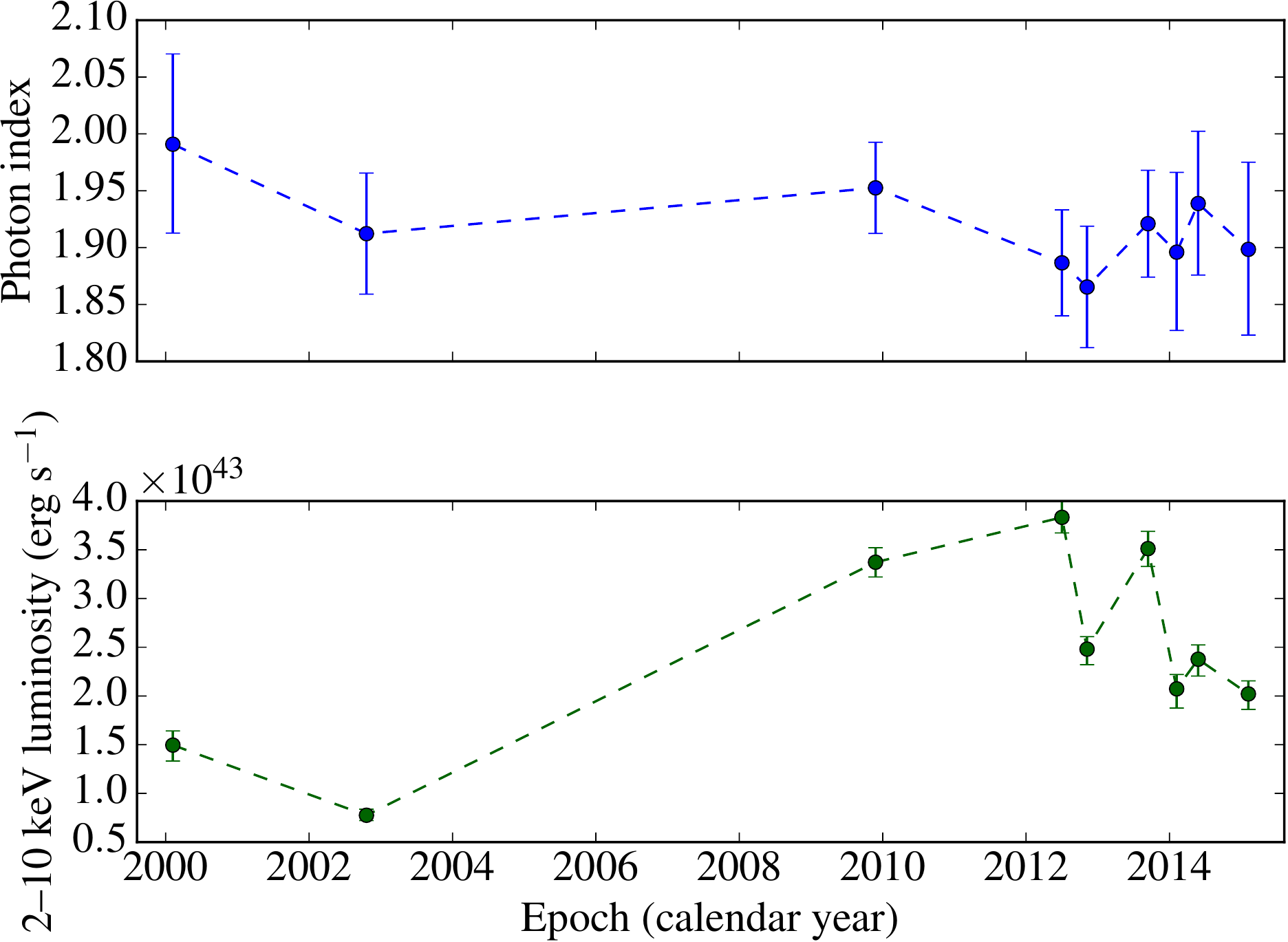}
\caption{The best-fitting photon indices and luminosities for
  a single power-law model of the AGN as a function of observing date.}
\label{fig:agn}
\end{figure}

The emission from the AGN is strongly piled up in the {\it Chandra}
observations and so our data are not particularly useful for studying
it in detail. However, we can obtain some information about AGN
variability over the period of the observations.

To deal with the effects of pileup we extracted spectra from annular
regions (as used by e.g.\ \citealt{Evans+04}) covering the wings of
the AGN point spread function (PSF). The annulus had an inner radius
of 6.3 pixels and an outer radius of 29 pixels -- the inner radius
excludes all regions of the AGN emission that are piled up at more
than the 1 per cent level at any epoch, while the relatively small
outer radius was chosen because the observations of epoch 2 have the
core close to the gap between ACIS-S and ACIS-I chips. For the same
reason, the background was taken from a circular region displaced
$\sim 1$ arcmin to the NW. The ancillary response files (ARFs) for the
spectra of the annular regions were then corrected for the
energy-dependent missing count fraction using point-spread functions
generated using the ray-tracing tool {\sc saotrace} and the detector
simulator {\sc marx} in the manner described by \cite{Mingo+11}, using
the satellite aspect solution appropriate for each observation so that
the simulated and real data are as close a match as possible. (Note
that the spectrum input to {\sc saotrace} does not affect the
correction factor, which is just derived from the ratio of the counts
in the annulus to the total counts in the PSF as a function of energy:
a flat input spectrum was used to achieve constant signal to noise in
the corrections.) The correction factors calculated are quite large
for soft X-rays, $\sim 40$ between 0.4 and 2.0 keV, but fall
significantly towards higher energies as expected.

We then fitted a single unabsorbed model\footnote{In models with an
  additional component of absorption at the redshift of Pic A, the
  fitted $N_{\rm H}$ is consistent with zero at all epochs, and a
  $3\sigma$ upper limit is typically 1-2 $\times 10^{20}$ cm$^{-2}$,
  i.e. significantly less than the Galactic column towards the AGN.
  Pic A, unlike some other broad-line radio galaxies, appears in our
  data to be a genuine `weak quasar' with an unobscured line of sight
  to the accretion disk. \cite{Sambruna+99} measured a slight excess
  absorption over the Galactic value in their {\it ASCA} observations,
  but {\it Chandra} has much better soft sensitivity and the {\it XMM}
  data also imply no excess absorption, so we believe our constraint
  to be more robust.} to the spectra for each epoch, obtaining the
results plotted in Fig.\ \ref{fig:agn}, where the 2-10 keV fluxes were
obtained using the {\sc sherpa} {\it sample\_flux} command. All fits
of this model were good, with reduced $\chi^2 \sim 1$. Unsurprisingly,
our results show that the AGN has varied significantly in total
luminosity over the period of our observations, though any variation
in photon index is much less prominent. In Table
\ref{table:litspectrum} we list some core photon indices and
luminosities from the literature and from the available archival {\it
  XMM} data, which show that the luminosities and photon indices we
obtain are very similar to those found in earlier work. In particular,
the {\it XMM} data confirm the low luminosity seen by {\it Chandra} in
the early 2000s but suggest that the source had returned to more
typical luminosities by 2005.

The individual epochs from the annulus observations are not sensitive
enough to search for Fe K$\alpha$ emission, but when we combine the
corrected annulus data and fit with a single power-law model (Table \ref{tab:spectra}) a narrow
feature around 6 keV is seen, which can be fitted with a Gaussian with
peak rest-frame energy $6.36 \pm 0.02$ keV, $\sigma = 50$ eV (fixed)
and equivalent width $330^{+30}_{-90}$ eV. It is possible that this
feature is itself variable (which would explain the discrepancy
between our equivalent width and the upper limit set by
\cite{Sambruna+99}) but our data are not good enough to test this
model further.

\begin{table*}
\caption{Literature/archive luminosities and photon indices for the AGN of
  Pictor A}
\label{table:litspectrum}
\begin{tabular}{llrrr}
  \hline
  Date&Telescope&Reference&Luminosity&Photon index\\
  &(instrument)&&(2-10 keV, erg s$^{-1}$)&\\
  \hline
  1996 Nov 23&{\it ASCA}&3&$3 \times 10^{43}$&$1.80 \pm 0.02$\\
  1997 May 08&{\it RXTE} PCA/HEXTE&4&$6 \times 10^{43}$&$1.80 \pm 0.03$\\
  2001 Mar 17&{\it XMM} PN&1&$1.82 \times 10^{43}$&$1.77 \pm 0.01$\\
  2005 Jan 14&{\it XMM} PN+MOS&2&$2.86 \times 10^{43}$&$1.775 \pm 0.002$\\
  \hline
\end{tabular}
\vskip 10pt
\begin{minipage}{11cm}
References are (1) HC05 (data re-analysed for this paper) (2)
\cite{Migliori+07} (data re-analysed for this paper) (3)
\cite{Sambruna+99} (4) \cite{Eracleous+00}, corrected to modern
cosmology. Note that the {\it Rossi X-ray Timing Explorer} data used
by reference (4) would have included contributions from the jet, lobe
and hotspot regions.
\end{minipage}
\end{table*}

\subsection{The jet and counterjet}
\label{sec:res-jet}

\begin{figure*}
\includegraphics[width=1.0\linewidth]{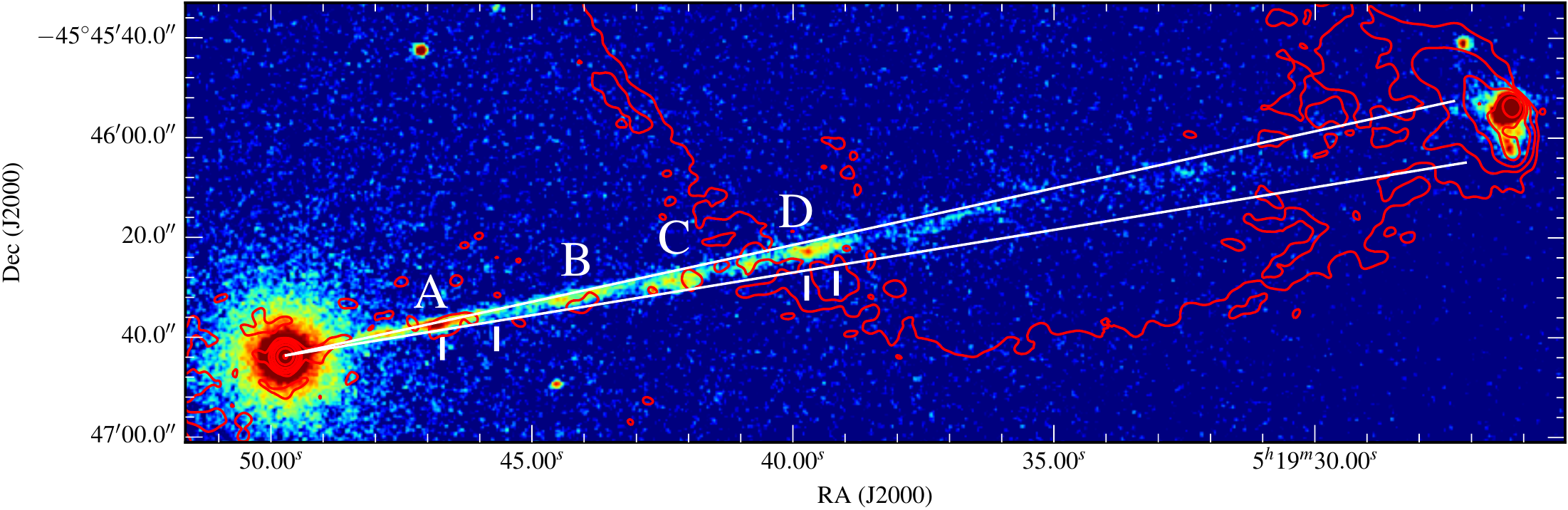}
\includegraphics[width=1.0\linewidth]{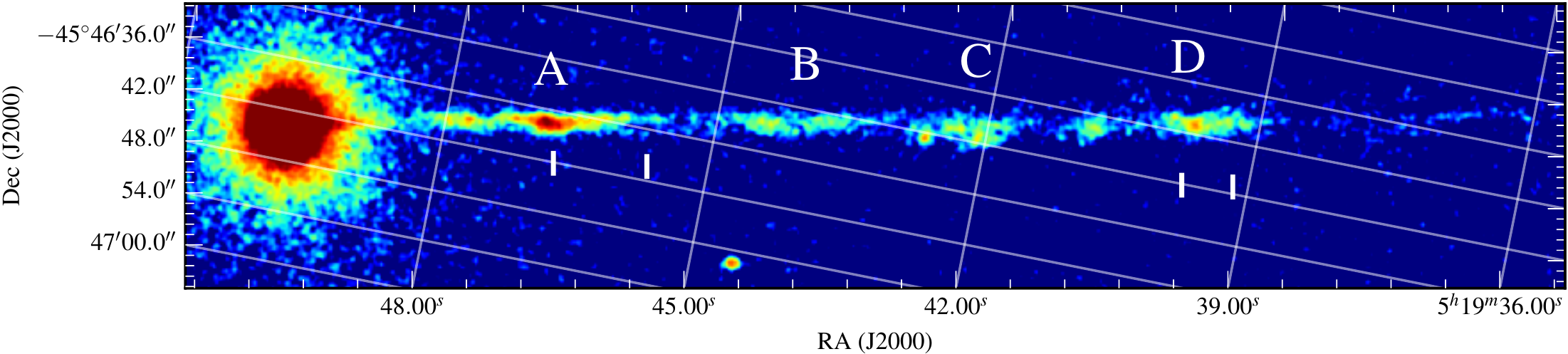}
\caption{X-ray emission from the jet. Top panel: raw counts in
  the 0.5-5.0 keV band, binned into 0.246-arcsec pixels and smoothed
  with a Gaussian with FWHM 0.58 arcsec. Superposed are contours of
  the ATCA 5-GHz image with a resolution of 2.2 arcsec (contours at 1,
  4, 16\dots mJy beam$^{-1}$). White diagonal lines indicate an
  opening angle of $3^\circ$ centred on the active nucleus. White
  vertical lines give the positions of candidate optical counterparts.
  Bright regions of the inner jet are labelled for reference in the
  text. Bottom panel: the same data, binned in 0.123-arcsec pixels and
smoothed with a Gaussian of FWHM 0.44 arcsec, rotated and zoomed in on
the inner jet.}
\label{fig:jetimage}
\end{figure*}

The radio jet of Pic A, first described by \cite{Perley+97}, is a very
faint, one-sided structure, hard in places to distinguish from
filamentary structure in the lobes. There is no detection of the jet
at wavelength between radio and X-ray, with the exception of a few
knots identified in the {\it HST} imaging by Gentry \etal\ (2015); for
example, it is not clearly visible in the available {\it Spitzer}
24-$\mu$m data. This makes the bright, knotty structure seen in the
X-ray all the more remarkable, as noted by W01. Many comparable
lobe-dominated, beamed systems with brighter radio jets show little or
no jet-related X-ray emission in {\it Chandra} images (e.g. 3C\,263,
\citealt{Hardcastle+02}; 3C\,47, \citealt{Hardcastle+04}). The
counterjet is not detected at any wavelength other than the X-ray;
again, continuous counterjet emission is unusual in FRIIs, although
there are several examples of knots from the counterjet side being
detected in narrow-line radio galaxies \citep{Kraft+05,Kataoka+08} and
there is a clear detection in at least one FRI \citep{Worrall+10}.

\subsubsection{Jet X-ray structure}
\label{sec:res-jetxray}

Fig.\ \ref{fig:jetimage} shows an image of the jet region with radio
contours overlaid. We begin by noting the following basic properties
of the X-ray jet:
\begin{itemize}
\item As stated above, the jet extends for all of the $\sim 4$ arcmin
  between the AGN and the hotspot. However, there is a very pronounced
  surface brightness change at 2 arcmin, just after the knot D
  indicated on Fig.\ \ref{fig:jetimage}. Little or no distinct compact
  structure is seen after this point. Hereafter we refer to the bright
  structure within 2 arcmin of the nucleus as the `inner jet' and the
  remainder as the `outer jet'.
\item The jet is quite clearly resolved transversely by {\it Chandra}
  over most of its length (conveniently placed point sources show the
  approximate size of the effective PSF at 1 and 4 arcmin from the
  nucleus).
\item The jet broadens with distance from the nucleus. The inner jet
  has an opening angle of roughly $3^\circ$, which, remarkably, is
  also the angle subtended by the X-ray hotspot at the AGN. It is hard
  to say whether the outer jet has the same opening angle, but
  certainly most of its emission is contained within boundary lines
  defined by the inner jet (Fig.\ \ref{fig:jetimage}).
\item There is strong variation in the surface brightness of the inner
  jet with distance from the nucleus, with particularly bright regions
  (labelled as `knots' A,B,C,D) at around 30, 60, 80 and 105 arcsec from
  the nucleus; the quasi-periodic spacing of these `knots' is striking. However, there are no locations where the surface
  brightness convincingly drops to zero. There is also some indication
  that the jet is not uniform transversely, in the sense that the
  brightest regions are displaced to one or the other side of the
  envelope defined by the diffuse emission (Fig.\ \ref{fig:jetimage}).
\item Although there are radio detections of the brightest X-ray
  features, the radio knots are not particularly well aligned with the
  X-ray features, and certainly do not match them morphologically.
  (However, we caution that the radio data are dynamic-range limited
  around the bright core, confused by structure in the lobes, and of
  intrinsically lower resolution than the X-ray data, so a detailed
  comparison is difficult.)
\end{itemize}

\subsubsection{Jet X-ray and broad-band spectrum}
\label{sec:res-jetspectrum}

\begin{figure}
\includegraphics[width=1.0\linewidth]{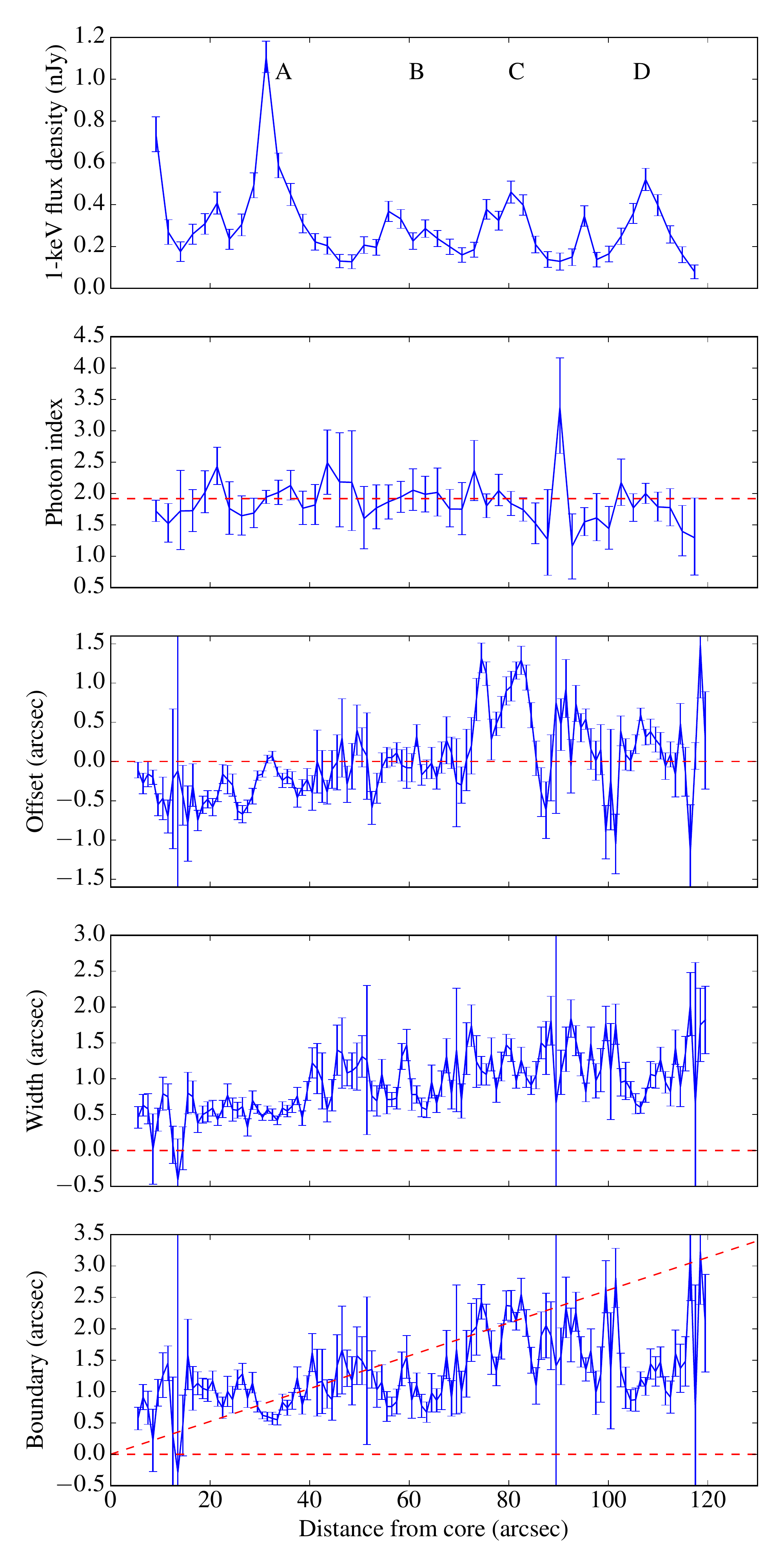}
\caption{Profiles of various quantities along the inner, bright part
  of the jet. First and second panels: flux density and photon index
  from spectral fits to the jets. The approximate positions of the
  brightness peaks in the jet are labelled in the first panel, and the
  best-fitting photon index for the whole inner jet is plotted as a
  red dashed line. Third panel: transverse offsets of the centroid of
  the jet from the mid-line, indicated by the red dashed line.
  Negative offsets are in a counterclockwise (roughly northern) sense,
  positive ones in a clockwise (southern) one. Fourth
  panel: the deconvolved width ($\sigma$) of the Gaussians fitted to the
  transverse profile. Fifth panel: the sum of the Gaussian width and
  the absolute value of the offset, giving an indication (since
  $\sigma$ is approximately the half-width at half-maximum) of the
  location of the outer envelope of the jet emission. The sloping red
  dashed line corresponds to a jet opening angle of $3^\circ$. Note
  that the 1-arcsec widths of the slices used in panels 3-5 means that
adjacent data points are not completely independent. See the text for
more details on the construction of the profiles.}
\label{fig:profile}
\end{figure}

We initially extracted spectra (Table \ref{tab:spectra}) for the inner and outer jet regions
separately, using rectangular extraction regions with adjacent
identical background regions (which account for lobe emission
  adjacent to the jet) and combining data from all observations
as discussed above. The 1-keV flux densities of these regions are
quite different ($11.7 \pm 0.2$ nJy versus $2.9 \pm 0.2$
nJy) but the photon indices are consistent (respectively $1.92 \pm
0.03$ -- note that the error here is probably underestimated because
of calibration uncertainties -- and $1.96 \pm 0.09$). Thus there is no
evidence for differences in the emission mechanisms in the two parts
of the jet.

We next divided the inner jet into small adjacent rectangular regions
with a length of 5 pixels (2.46 arcsec) and width 16 pixels (7.8
arcsec). These regions are wide enough that we should be looking at
resolved regions of the jet and that variations between the PSFs of
different observations should have little effect. Starting at 8 arcsec
from the core, we extracted spectra for each region, 47 in total over
the full extent of the inner jet. The results are shown in
Fig.\ \ref{fig:profile}. We see that there is no evidence for
significant changes in the jet photon index as a function of length.
Only one region, a region of low surface brightness in between knots C
and D, shows weak evidence for a significantly different X-ray
spectral index, unlike the case in the best-studied FRI jet, that of
Cen A, where clear systematic trends in the jet photon index as a
function of position are seen
\citep{Hardcastle+07}.

\subsubsection{Jet emission profile}
\label{sec:res-jet-profile}

To quantify the structure seen in the images of the jet
(Fig.\ \ref{fig:jetimage}) we next divided the jet up into finer
regions (1 arcsec long by 10 arcsec across the jet) and fitted a model
consisting of a flat background and a Gaussian to the events of each
slice, using a likelihood method with Poisson statistics. We fitted
only to slices detected at better than the $2\sigma$ level. The width
($\sigma$) and position of the Gaussian (in terms of its angular
offset from the mid-line of the jet) were free to vary, as was the
background level, which was additionally constrained by fitting to
adjacent 10-arcsec regions containing no jet emission. The width of
the Gaussian was combined with the expected $\sigma = 0.34$ arcsec of
the un-broadened PSF to give a rough deconvolution of PSF effects. In
general, we found that a transverse Gaussian gave an acceptable fit to
the profile slices; there was no evidence for significant
edge-brightening. However, there are clear variations in the size and
offset of the Gaussians along the jet, as shown in Fig.
\ref{fig:profile}. The jet gets systematically wider with length, and
we see that the inner knots are systematically displaced to the N
while knot C is systematically S of the centre-line (which is
approximately the line between the core and the brightest part of the
hotspot). The outer envelope of the jet (roughly estimated as the sum
of the Gaussian width and its offset) also gets larger with distance
from the core, and it can be seen that at large distances the envelope
is roughly consistent with a constant opening angle around $3^\circ$.
At distances $\la 20$ arcsec from the core the jet appears to be
slightly resolved with a constant Gaussian width of about 0.5 arcsec.

The analysis we have carried out is very similar in intention and
methods to the analysis of the radio data for the straight jets of
four powerful lobe-dominated quasars by \cite{Bridle+94}, so it is
interesting to compare our results with theirs. Like us, they see a
roughly linear increase of jet width with length in two sources with
very well-defined straight jets (3C\,175 and 3C\,334). The opening
angles in these jets are similar to those seen in Pic A
(2--3$^\circ$). Having said that, the two other quasars they study in
detail show little or no trend with distance, and there is some
evidence that 3C\,334 recollimates at large distances, so it is not
clear that all these jets can be expanding freely over their length.
We return to the implications of the apparent constant opening angle
in Pic A below, Section \ref{sec:ratio}.

\subsubsection{Jet variability}
\label{sec:res-jetvar}

\begin{figure}
\includegraphics[width=1.0\linewidth]{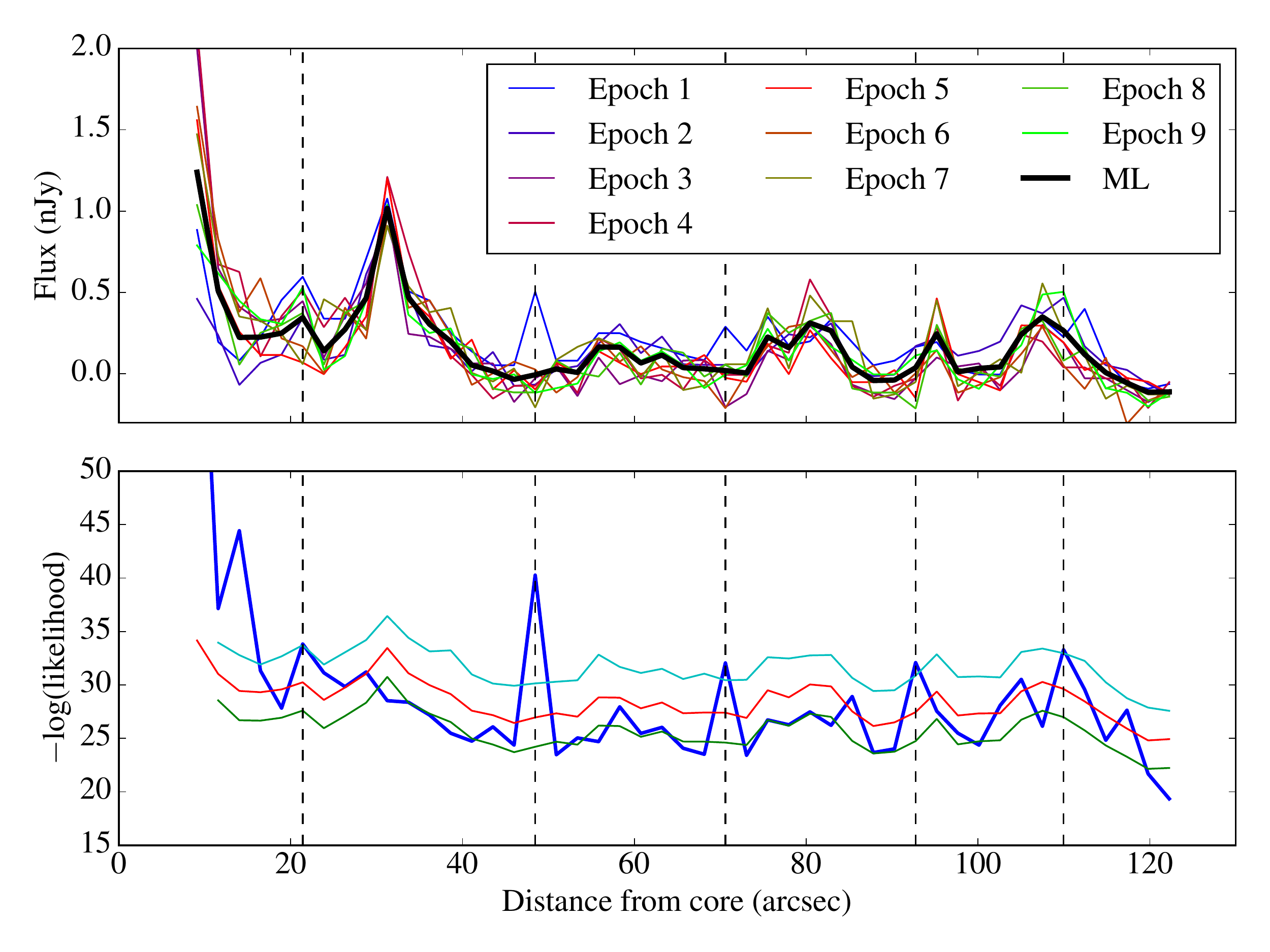}
\caption{Profile of the flux density as a function of epoch (top) and
  the negative log-likelihood of the best-fitting constant-flux model
  (bottom) along the inner jet. See text for a description of the
  points. The top panel is colour-coded by observing epoch, with the
  thick dark blue line giving the maximum-likelihood flux density on
  the assumption of a constant flux across all epochs as a function of
  the position of the extraction region: error bars are not plotted
  for clarity. The bottom panel shows the log-likelihood for the fit
  at that position as the thick blue line and the expected value and
  90 per cent and 99 per cent upper bounds as thin green, red and cyan
  lines respectively.}
\label{fig:variability}
\end{figure}

To assess the level of variability in the jet we used the same regions
as in the previous section, but now divided into the 9 epochs of
observation listed in Table \ref{tab:obs}. There are not enough counts
in each region after this division to allow fitting of models to the
  extracted X-ray spectra as a function of time in {\sc xspec} or {\sc
    sherpa}, even
with fixed photon index; moreover, the errors on the counts in
individual regions are quite high if the adjacent local background
regions are used. We therefore used the following procedure:
\begin{itemize}
\item We determined for each epoch a background level in counts by
  amalgamating all background regions at more than 30 arcsec from the
  core, having verified that there are no systematic trends in the
  background level as a function of distance from the core at any
  epoch. The statistical errors on these background levels are
  negligible compared to the Poisson errors on counts in individual
  regions.
\item We computed the conversion factor between 1-keV flux density and
  0.4-7.0 keV counts for each region and epoch, using the response
  files generated to measure the photon indices shown in
  Fig.\ \ref{fig:profile} and a fixed photon index corresponding to
  the best-fitting value for the inner jet. These conversion factors
  generally vary little with distance along the jet for a given epoch,
  and are featureless apart from the effect of CCD node boundaries, but of course vary quite
  significantly between epochs. The conversion factors allow us to
  plot the best estimate of the flux density profile at each epoch
  (Fig.\ \ref{fig:variability}).
\item With these conversion factors and the background levels we can
  compute the maximum-likelihood flux density for each region on the
  assumption of a constant flux at all epochs: essentially this is the
  same approach as used in the Cash statistic for model fitting
  \citep{Cash79}. Low values of the maximum likelihood (equivalently,
  high values of the natural log of the reciprocal of the likelihood,
  plotted in Fig.\ \ref{fig:variability}) imply a poor fit of a
  constant-flux model.
\end{itemize}

\begin{figure}
\includegraphics[width=1.0\linewidth]{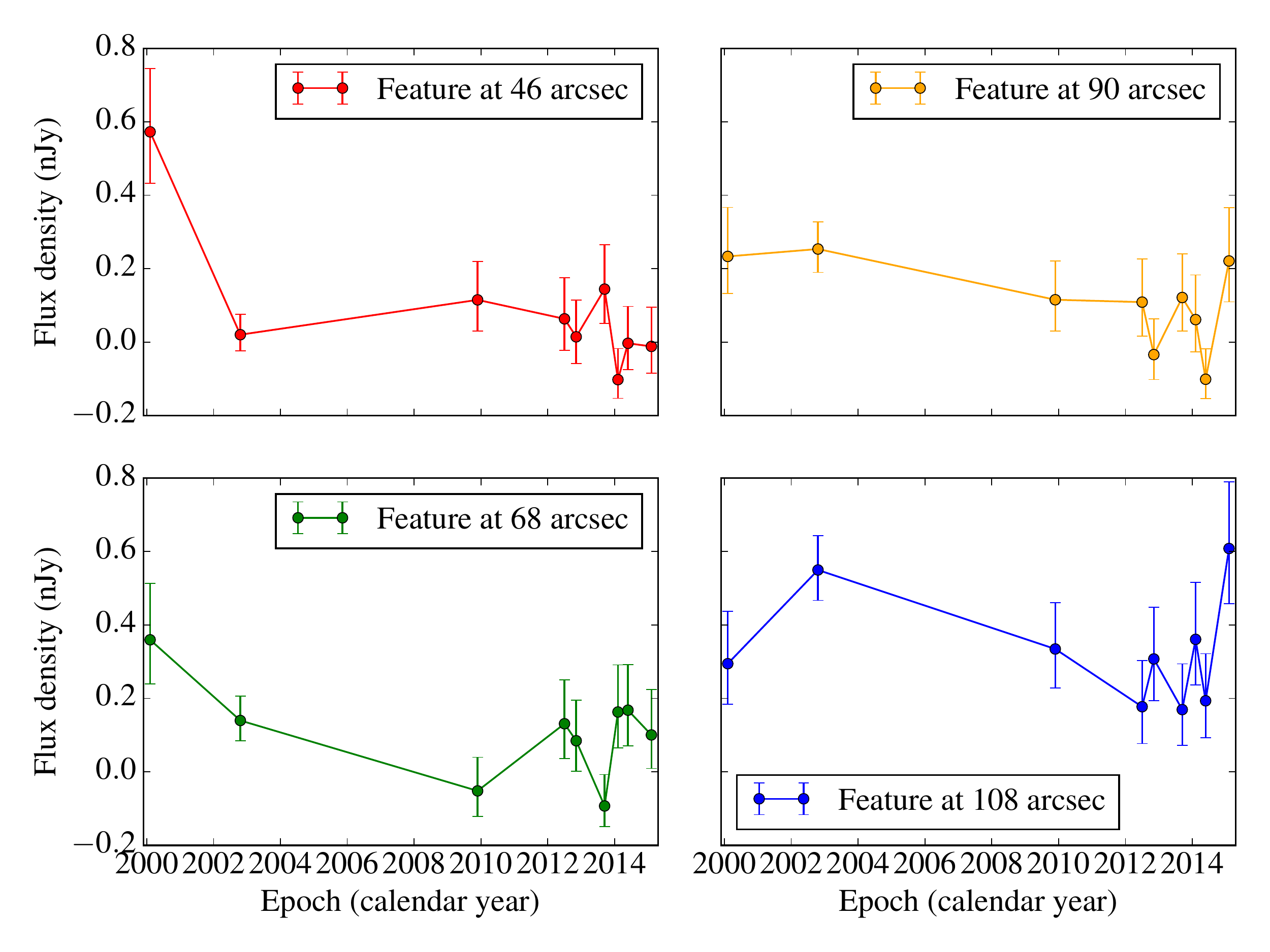}
\caption{Light curves for potentially variable features identified by
  the maximum-likelihood analysis. Error bars are plotted using the
  methods of \protect\cite{Gehrels86}.}
\label{fig:jet-lc}
\end{figure}

From Fig.\ \ref{fig:variability} it can be seen that there are several
peaks in the profile of the fitting statistic, corresponding to the
flare reported by M10 at 48 arcsec and the possible feature at 70
arcsec as well as to other locations. To assess the significance of
these variations we determined the expected log-likelihood if the data
were in fact consistent with the best-fitting constant-flux model
(green line on Fig.\ \ref{fig:variability}): we used Monte Carlo
methods to do this, taking account of the Poisson errors on the counts
in each bin, though it could be done analytically. The problem
of significance now in principle reduces to a classical likelihood
ratio test, but since there are few counts per bin we chose not to
make use of the fact that the asymptotic distribution of the log of
the likelihood ratio is the $\chi^2$ distribution: instead we computed
confidence levels at the 90 and 99 per cent levels by running the
Monte Carlo simulations many times to assess the distribution of the
log-likelihood per bin. Setting aside the inner part of the jet, where
the apparent variability is probably dominated by the AGN (we have
made no attempt to subtract the wings of the PSF), we see peaks at
better than 99 per cent confidence (before accounting for trials) at
22, 48, 70, 92 and 110 arcsec; by far the most significant feature is
the original flare of M10 at 48 arcsec (34 kpc in projection). Given
that the variability of the core might still affect the inner jet at
22 arcsec -- the plot shows a systematic downward trend of the maximum
likelihood inside $\sim 40$ arcsec, which is plausibly due to core
contamination -- we suggest that only the regions beyond this point should
be taken at all seriously: it is notable that three out of four of
the potentially variable sources beyond 22 arcsec lie in inter-knot
regions of the jet (between A and B, B and C, and C and D
respectively). Light curves for these variable regions are shown in Fig.
\ref{fig:jet-lc}.

It is clear that no new flares comparable to the one reported by M10
have taken place, though we may be seeing lower-level variability in
other parts of the jet. Of course, we cannot claim a 99 per cent
confidence detection of variability in any other individual region
because we have carried out $\sim 40$ independent trials, which
reduces the individual significance, but the fact that we have more
than one region above the 99 per cent confidence limit increases the
probability that at least some of them are real. Further Monte Carlo
simulation shows that the expected average number of spurious
`detections' over the whole jet beyond 20 arcsec at the 99 per cent
confidence level derived as above, on the null hypothesis of no actual
variability, is 0.46 (very similar to the level expected on a naive
analysis), compared to the 5 detections reported above; there is a 37
per cent chance that one such detection is spurious, an 8 per cent
chance that two are, and only a 1 per cent chance that three are
spurious, so it seems very likely that some of the newly detected
variable regions are real, though we cannot say which. We can rule out
the possibility that the apparent variability is produced by some
global {\it Chandra} calibration error, since this would be expected
to produce correlated variability between points at the same epoch,
which is not observed; as noted above, there is no evidence for
small-scale features in the point-to-point count-to-flux conversion
factors.

We comment on the implications of the results on jet variability in
Section \ref{sec:discussion-jetvariability}.

\subsubsection{The counterjet}

The counterjet is much fainter than the jet and is detected at high
significance for the first time in these observations. It is not
visible very close to the nucleus, and merges into the diffuse
emission associated with the E hotspot. It is notable that it does not
align with the brightest radio structures in that hotspot,
though it does point towards a bright X-ray feature (see below,
Section \ref{sec:ehotspot}).

We extracted a spectrum for the detectable part of the counterjet,
using a rectangular region of length 113 arcsec and height 16 arcsec
centered in the E lobe and avoiding the diffuse emission around the E
hotspot, again with local background subtraction. We find a 1-keV flux
density of $1.6 \pm 0.3$ nJy and a photon index of $1.7 \pm 0.3$. Thus
we see no evidence from the X-ray spectrum that the jet and counterjet
have different emission mechanisms.

We carried out the same profiling analysis as described in Section
\ref{sec:res-jet-profile} for the counterjet, but most regions were
too faint to be fitted even with large (5-arcsec) regions. There is
some evidence that the counterjet is slightly broader at larger
distances from the core, but the error bars are large.

\subsection{The western hotspot}
\label{sec:whotspot}

\begin{figure}
\includegraphics[width=1.0\linewidth]{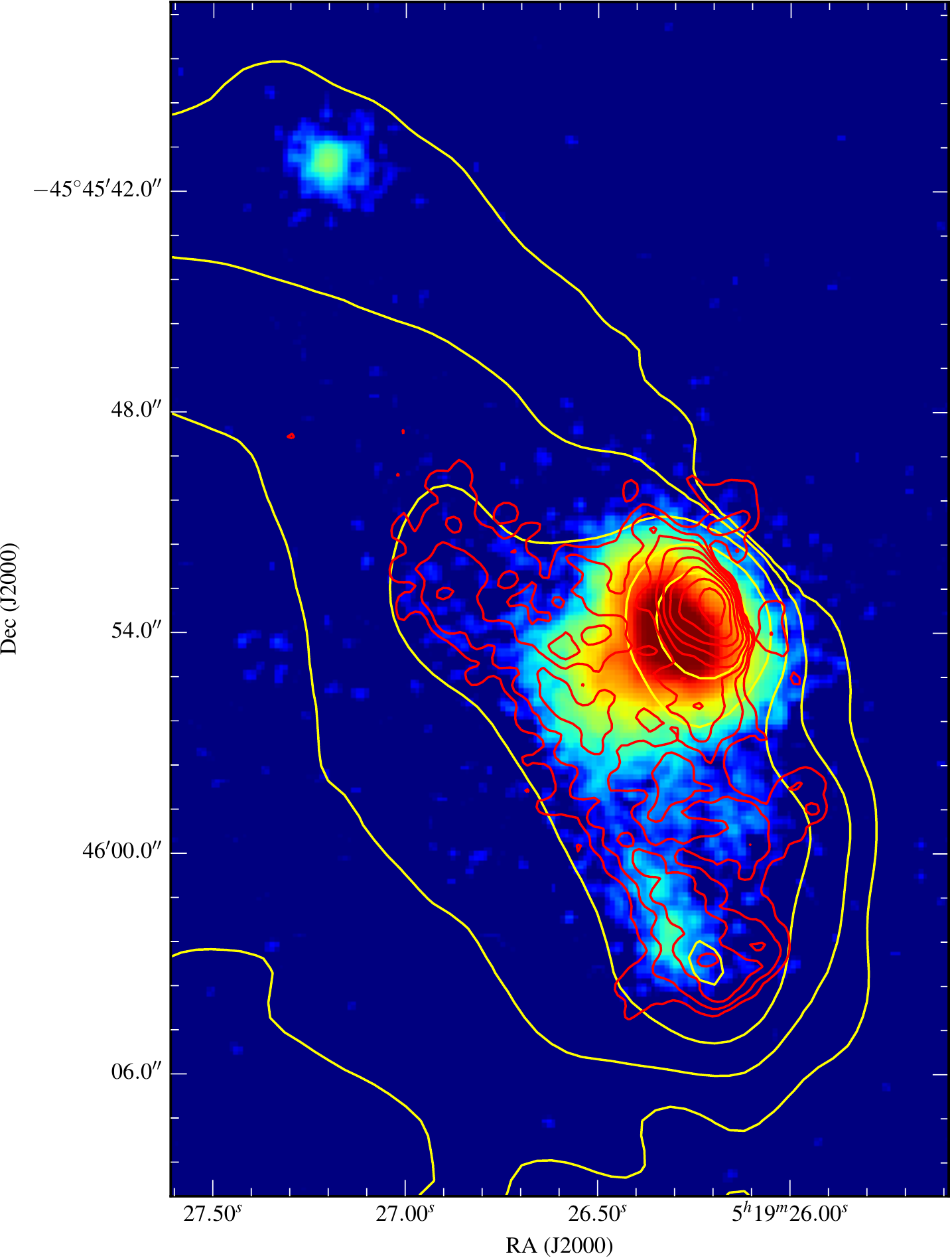}
\caption{The W hotspot. Logarithmic colour scale shows counts in the
  0.5-5.0 keV passband, binned to pixels of 0.123 arcsec on a side and
  smoothed with a Gaussian of $\sigma = 1$ pixel to give an effective
  resolution of $\sim 0.7$ arcsec. Overlaid are
  contours from the 5-GHz ATCA map with 1.7-arcsec resolution at 2,8,32\dots
  mJy beam$^{-1}$ (yellow) and contours from the 15-GHz VLA map of
  \protect\cite{Perley+97} with 0.5-arcsec resolution at 1,2,4\dots mJy
  beam$^{-1}$ (red). A presumably unrelated X-ray point source to the N
  of the image gives an indication of the effective PSF of the stacked {\it
    Chandra} data.}
\label{fig:whotspot}
\end{figure}

Fig.\ \ref{fig:whotspot} shows an overlay of the {\it Chandra} and
radio images of the W hotspot.

We begin by noting that the high count rate in the hotspot is not
necessarily wholly beneficial -- for the count rates in the epoch 2
observations, where the hotspot is at the aim point, there is some
possibility of pileup given the count rates of $\sim 0.2$ s$^{-1}$. We
see no evidence of significant grade migration in the data for these
epochs, probably because the hotspot is resolved (see below) and so do
not attempt to correct for pileup in any way. The possibility of a
small pileup effect (leading to a harder spectrum) should be borne in
mind in the interpretation of our results.

W01 remarked on the strong similarity between the radio through
optical morphology as seen by \cite{Perley+97} and the X-ray, and
these deeper data confirm that, though they also point to some
interesting differences. The most striking is a clear offset of around
1 arcsec (0.7 kpc in projection) between the peak radio and X-ray
positions of the hotspot, in the sense that the X-ray emission is
recessed along the presumed jet direction; this offset is visible when
comparing to both ATCA and VLA data (which, by contrast, appear well
aligned with each other) and is clearly real so long as our astrometry
is reliable (see above, Section \ref{sec:obs-atca}). Similarly, the
extension of the hotspot to the SE is not so prominent in the radio or
optical data, and the surface brightness distribution of X-ray and
15-GHz radio is rather different in the `bar' to the E of the compact
hotspot. The X-ray bright part of this bar region, directly S of the
peak X-ray emission, is consistent with being unresolved transversely
by {\it Chandra}.

The integrated spectrum of the entire hotspot region, using a circular
aperture of radius 10 arcsec which encompasses all the emission, can
be fitted with a power law with $\Gamma = 1.94 \pm 0.01$ -- note the
similarity to the jet photon index -- and total 1-keV flux density
$90.5 \pm 0.5$ nJy. However, the fit is not particularly good (Table
\ref{tab:spectra}). A better fit is obtained with a broken power law, with
a break energy of $2.1 \pm 0.2$ keV and photon indices below and above
the break of $1.86 \pm 0.02$ and $2.16_{-0.04}^{+0.06}$ respectively,
and an almost identical 1-keV flux density\footnote{Note that a
  similar broken power-law model is an acceptable fit to the jet, though the data
  quality in the jet are not sufficient to constrain the break energy
  or to distinguish between this model and a single power low.}. A pure thermal model for
the hotspot is conclusively ruled out, with $\chi^2 = 567/303$ even
when the metal abundance is (unrealistically) allowed to go to
zero. A model with a power law and APEC thermal model (with abundance
fixed to 0.3 solar, since otherwise abundance and power-law normalization
are degenerate) is a less good fit than the broken power law (Table
\ref{tab:spectra}).

To investigate whether the broken power-law best fit is the result of
the superposition of two different spectra, we divided the hotspot
into non-overlapping `compact' and `bar' components, where the
`compact' region is an ellipse around the brightest part of the X-ray
hotspot and the `bar' region is a rotated rectangle encompassing the
linear structure seen in radio emission to the E. Interestingly, these
two regions do have different photon indices on a single power-law fit
($1.97 \pm 0.01$ and $1.83 \pm 0.03$ for the compact and bar regions
respectively: in {\it comparing} the two photon indices we may neglect the
calibration uncertainties since the two regions have essentially the
same calibration applied). However, the single power-law model remains
a poor fit to the compact region and once again a broken power law is
better (Table \ref{tab:spectra}), with $E_{\rm break} =
2.14_{-0.14}^{+0.23}$, $\Gamma_{\rm low} = 1.87 \pm 0.02$ and
$\Gamma_{\rm high} = 2.23_{-0.04}^{+0.07}$. This same model, with only
normalization allowed to vary, is an acceptable fit ($\chi^2 =
170.0/145$) to the bar region, although the single power-law fit is
better, so there is no strong evidence for spectral differences in the
two components: in any case, the steepening of the X-ray spectrum
appears to be intrinsic to the compact region of the hotspot.

\begin{figure}
\includegraphics[width=1.0\linewidth]{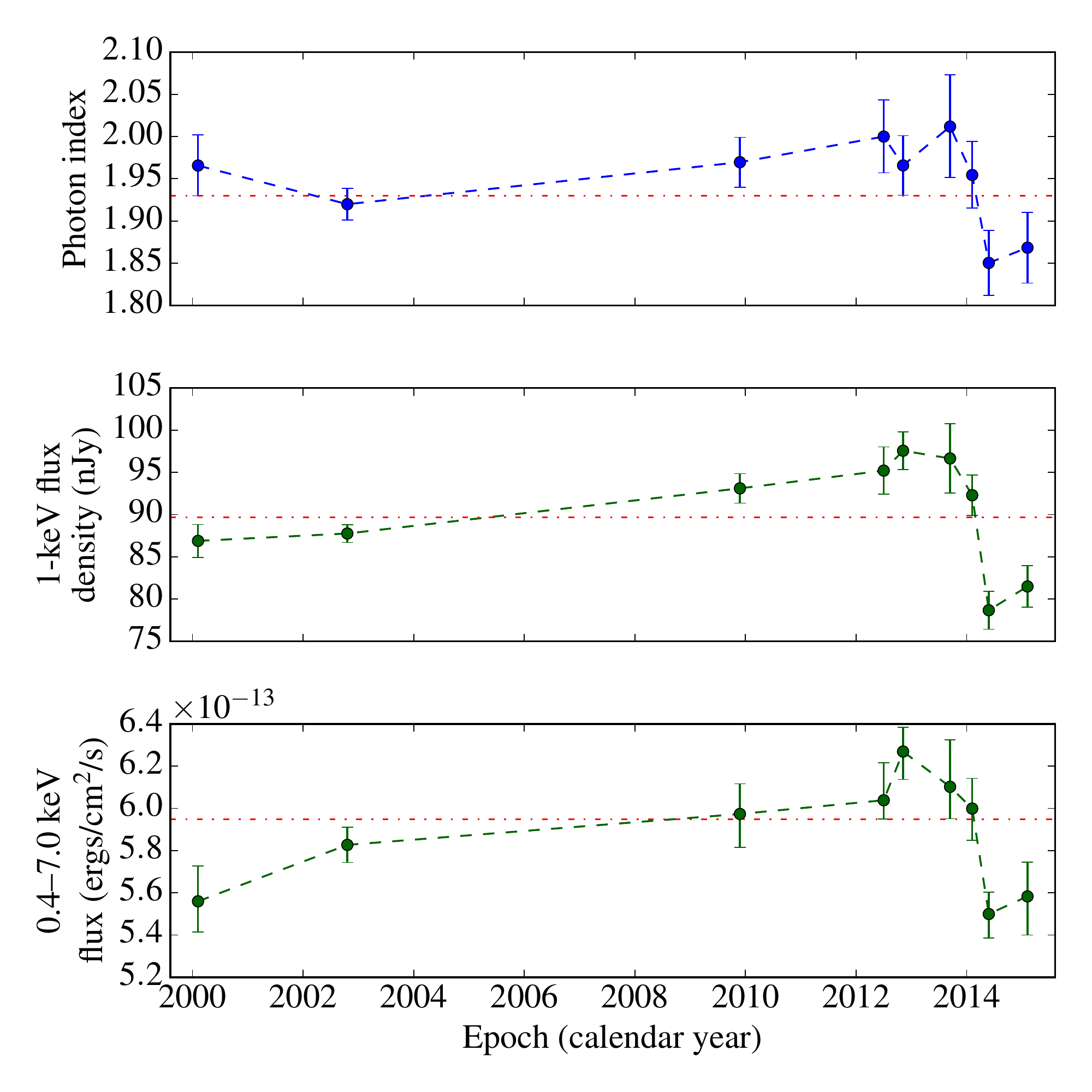}
\caption{The best-fitting photon indices, 1-keV flux densities and
  total flux in the {\it Chandra} band for
  a single power-law model of the W hotspot as a function of observing
  date. Red dashed lines show the values derived from a joint fit to
  the data, effectively a weighted mean for all the observations.}
\label{fig:hotspot-variability}
\end{figure}

We searched for variability in the hotspot by fitting single power-law
models to the datasets from the individual epochs (using the single
large extraction region) and comparing the normalization and photon
index (Fig.\ \ref{fig:hotspot-variability}). Remarkably, there is some
evidence for variations in 1-keV flux density at the 5-10 per cent
level on our observing timescale, which would imply, if real, that a
significant fraction of the X-ray flux from the hotspot is generated
in compact regions with sizes of order pc or even less. These flux
density variations are reflected in variations in the total flux in
the {\it Chandra} band, showing that they are not simply the result of
the correlated variations in photon index (errors plotted on the flux
curve take the variations of both parameters of the fits into
account). Particularly striking is the drop in flux or flux density at
the 10 per cent level between epochs 7 and 8 (a timescale of only
three months). The data are not good enough to fit broken power laws
to the individual datasets, and so it is unclear whether the
best-fitting broken-power-law spectrum for the integrated hotspot
emission is in fact simply a reflection of this apparent temporal
variability. (The associated variations in spectral index are only
marginally significant, particularly if calibration uncertainty is
taken into account, and so we do not attempt to interpret them; in
particular the apparently flat spectrum in epoch 2 with respect to
other epochs might conceivably be an effect of pileup, as noted above,
and so should not be taken too seriously.) In epoch 8 the hotspot was
very close to a chip gap on the detector and, while the weighted
responses that we use should take account of that, the spectrum is
less trustworthy than at other epochs: however, as an essentially
identical fit is found to the data for epoch 9, where the hotspot is
in the centre of the ACIS-S3 chip, we are confident that the large
apparent drop in flux is not an instrumental artefact. We cannot, of
course, rule out some large and otherwise unknown error in recent
calibration files, but it is important to note that the AGN does not
show the same time variation between these two epochs (Section
\ref{sec:AGN}). On the assumption that we are seeing a real physical
effect, we discuss the implications of hotspot variability in Section
\ref{sec:discussion-hsvariability}.

\subsection{The eastern hotspot}
\label{sec:ehotspot}

\begin{figure}
\includegraphics[width=1.0\linewidth]{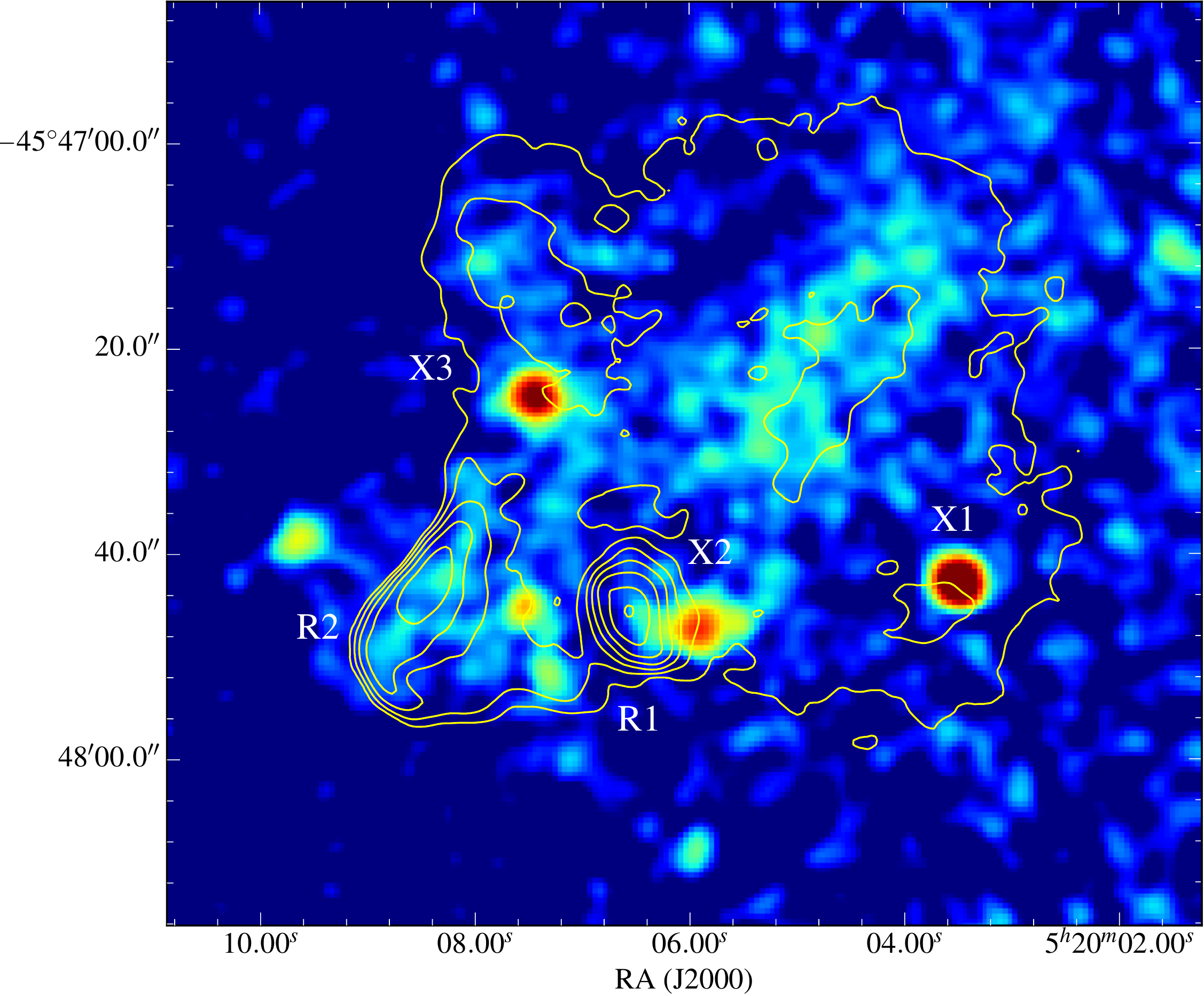}
\caption{The E hotspot. Logarithmic colour scale shows a fluxed image in the
  0.5-5.0 keV passband, binned to pixels of 0.492 arcsec on a side and
  smoothed with a Gaussian of $\sigma = 2$ pixel. Overlaid are
  contours from the 5-GHz ATCA map with 1.7-arcsec resolution at $1.5
  \times (1,2,4,\dots)$ mJy beam$^{-1}$.}
\label{fig:ehotspot}
\end{figure}

The E hotspot is a much more complex, and much fainter structure than
the W hotspot in both radio and X-ray, and accordingly we are more
limited in the investigations we can carry out. A radio/X-ray overlay
is shown in Fig.\ \ref{fig:ehotspot}. It can be seen that essentially
the whole region of excess radio surface brightness (the `hotspot
complex' in the terminology of \cite{Leahy+97}) is also enhanced in
the X-ray, though there is not a simple relation between diffuse radio
and X-ray emission (e.g. the peak of the diffuse emission in the
centre of Fig.\ \ref{fig:ehotspot} is not in the same place in radio
and X-ray). The relationship between the three bright compact X-ray
sources (labelled X1, X2, X3 in the figure) and the two compact radio
hotspots (R1, R2) is similarly unclear. X2 is clearly partly resolved in the
full-resolution {\it Chandra} image, which makes it less likely to be
a background source: if it is physically associated with the more
compact radio component, R1, then the offset of 7 arcsec (5 kpc)
between the X-ray and radio peaks is significant. X1 and X3 may be
background sources, but both lie at the edge of real, diffuse radio
features visible in the contour map, and neither has an optical
counterpart on Digital Sky Survey images. The counterjet, where last
visible, points directly towards X3. There is no compact
X-ray source associated with R2, but it is clearly associated with
enhanced X-ray emission.

The best-fitting power-law model applied to the entire elliptical
hotspot region gives a relatively flat photon index with $\Gamma = 1.8
\pm 0.1$. (The background region is a concentric ellipse, so
background from the lobes is at least partially subtracted from this
flux density value.) If we exclude X1 and X3, we obtain a consistent
$\Gamma = 1.8 \pm 0.1$ (Table \ref{tab:spectra}). Consistency of the
photon index with that of the jet or W hotspot region is not ruled out
at a high confidence level.

Finally, we draw attention to the apparent extension of the X-ray
emission to the E and S of the sharp boundary of the radio emission at
hotspot R2 (and therefore with no radio counterpart). This is
not seen at high significance -- the emission corresponds only to a
few tens of counts -- and the point source immediately to the NE of
R2, which contributes to it, is surely unrelated to the radio galaxy.
But it is possible that we are seeing here at a very low level shocked
emission from the thermal environment of the source. There are
insufficient counts to test this model spectrally, and no comparable
feature can be seen around the W hotspot.

\subsection{The lobes}
\label{sec:lobes}

\begin{figure}
\includegraphics[width=1.0\linewidth]{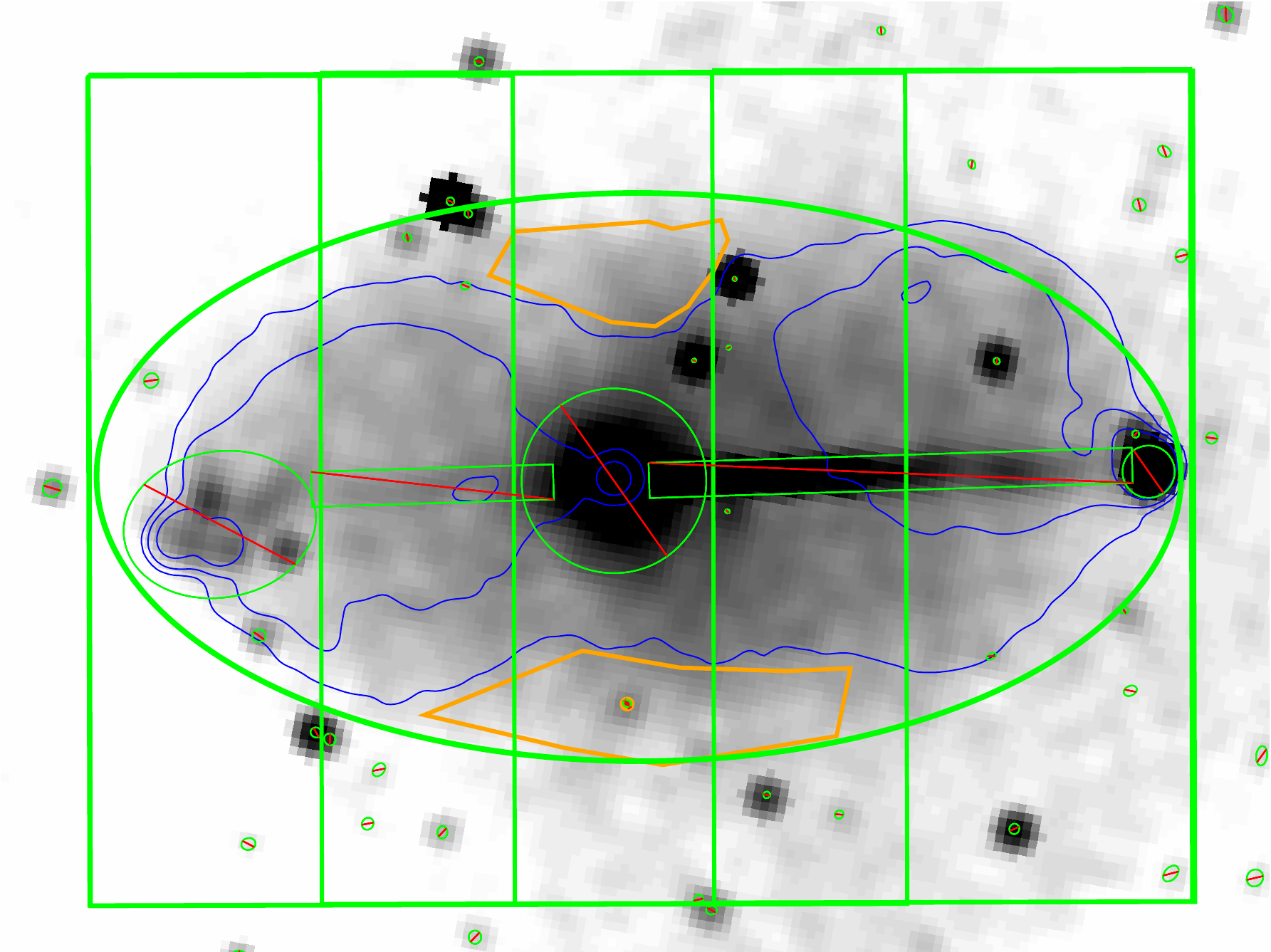}
\caption{The regions used for lobe spectral extraction. The greyscale
  is a binned, smoothed image of the 0.4-2.0 keV {\it Chandra} counts.
  The large green ellipse shows the basic lobe region: the rectangles
  indicate the sub-division into five smaller regions for spectral
  fitting and the orange polygons are the extraction regions for the
  emission outside the radio contours discussed in the text. Exclusion
  regions for core, jets, hotspots and background sources are shown
  defaced with red lines. Also plotted (in blue) are contours from the
  7.5-arcsec resolution 1.4-GHz radio map of \protect\cite{Perley+97}:
  contours are at 10, 40 and 160 mJy beam$^{-1}$.}
  \label{fig:lobe-regions}
\end{figure}

Because the lobes are significantly contaminated by scattered hard
emission from the PSF close to the nucleus, and this cannot be
corrected by local background subtraction, we restrict ourselves to
the energy range 0.4-2.0 keV in spectral fitting in this
section\footnote{This is more conservative than the approach used by
  HC05, since the AGN is substantially brighter relative to the lobes
  in the newer observations.}.

We initially carried out spectral fitting to the whole lobe region
(encompassing all of the E and W lobes with the exclusion of a
45-arcsec circle around the core and appropriate regions around the
jet and hotspot) and also to a sub-division of this large region into
five sub-regions in linear slices along the lobe (Fig.\ \ref{fig:lobe-regions}). The resulting spectrum for the whole lobe is flat
($\Gamma = 1.57 \pm 0.04$) and there is no evidence in the spectra of
the sub-regions for significant variation as a function of length
along the lobe, whether for physical reasons or as a result of
residual contamination by the AGN (Table \ref{tab:spectra}). The 1-keV
flux density and spectral index we obtain are in reasonable agreement
with those reported by HC05, who measured spectra and fluxes from the
two lobes separately; HC05's spectra are a little steeper, but it is
possible that this is a result of their use of the whole 0.4-7.0 keV
band for spectral fitting, as the inverse-Compton spectrum is expected
to steepen across the {\it Chandra} band.

A conspicuous feature of the X-ray `lobe' emission is that it extends
further than the radio contours at the centre of the source: this can
be seen in both Fig.\ \ref{fig:all} and Fig.\ \ref{fig:lobe-regions}.
We do not believe that this emission is residual scattered flux from
the nucleus, since, although the wings of the PSF are not negligible
in this region even in the 0.4-2.0 keV energy range, the predicted
surface brightness of emission from the {\sc saotrace}/{\sc marx}
simulations described in Section \ref{sec:AGN} at these radii would be
at least a factor 4-5 below what is observed. In an inverse-Compton
model, we would always expect radio emission at some level coincident
with the X-ray emission, leaving two possibilities: (1) this is
genuinely inverse-Compton emission from the lobes, and so there is
radio emission, but it is too faint and/or steep-spectrum to be
detected; or (2) we are seeing thermal emission from the otherwise
undetected hot gas halo around the lobes (which must be present at
some level to confine them, and which would be expected to be
particularly bright between the lobes). Possibility (1) cannot be
ruled out at this point: contours of the 330-MHz images of
\cite{Perley+97} do appear to include all the X-ray emission, but they
are much lower in resolution than any other map we have used here (the
resolution is $30 \times 6$ arcsec, with the 30-arcsec major axis
being in the N-S direction) and so do not provide strong constraints.
To investigate possibility (2) we extracted spectra for the regions
outside the lowest contour of the L-band image shown in
Fig.\ \ref{fig:lobe-regions}, and fitted them with thermal and
non-thermal (power-law) models. The results are inconclusive (Table
\ref{tab:spectra}): a power-law model is a good fit to the data but
with a rather steep photon index of $2.0 \pm 0.1$, a thermal (APEC)
model with abundance fixed to 0.3 solar fits somewhat more poorly than
the power law and gives an implausibly high temperature of $2.7 \pm
0.5$ keV, and when we fit a combination of the two, fixing the
power-law photon index to the value of 1.57 derived from the
whole-lobe region, the fit is dominated by the power-law component and
is no better than for the pure power-law model, though the derived
temperature is more reasonable for a poor environment. Similar results
are obtained from a powerlaw plus thermal fit to the middle lobe
region. While we cannot rule out the possibility of some soft thermal
emission, with a temperature consistent with being the environment of
the host elliptical, contributing to the observed X-rays in this
region, we see no compelling evidence that it is detected.
High-fidelity low-frequency radio maps will be needed to test
possibility (1) further.

HC05 have already discussed the evidence for large-scale variation in
the X-ray to radio ratio across the lobes, and we do not repeat their
analysis here. Fig.\ \ref{fig:all} already shows that any large-scale
surface brightness variation in the X-ray lobes is much smaller than
that in the 1.4-GHz radio emission. However, one thing that we can do
with the larger volume of data available to us is to study the
radio/X-ray ratio in a statistical way. As discussed in Section
\ref{sec:intro-lobes}, the objective here is to test models for the
origins of the `filaments' that appear to dominate small-scale surface
brightness variation in the radio lobes. In the extreme case in which
the variation in synchrotron emissivity that they imply is purely due
to variations in the normalization of the electron energy spectrum,
with a uniform magnetic field strength, then we would expect a
one-to-one relationship between the radio and X-ray emission. (This
model already seems to be ruled out by the observations of HC05, though we
comment on it more quantitatively below.) If, on the other hand, the
variation in synchrotron emissivity is only due to point-to-point
variations in magnetic field strength, with a uniform electron
population filling the lobes, then we would see a uniform
X-ray surface brightness (modulo line-of-sight depth effects) and thus
little correlation between the radio and X-ray emission. In between
these two extremes lie a range of models in which the local electron
energy spectrum normalization depends on magnetic field to some
extent.

To search for correlations between radio and X-ray we measure radio
flux densities, and X-ray fluxes, from as large a number of discrete
regions of the lobe as possible. Because we wish to search for
counterparts of the filamentary structures seen when the lobes are
well resolved, we use the highest-resolution radio map available to
use that does not resolve out lobe structure, the 7.5-arcsec
resolution 1.4-GHz map of \cite{Perley+97}
(Fig.\ \ref{fig:lobe-regions}). Ideally we would work at even lower
frequencies, since the electrons responsible for the observed
inverse-Compton emission emit at 20 MHz for a mean magnetic field
strength of 0.4 nT (Section \ref{sec:discussion-lobe}), but
high-resolution, high-fidelity images of the Pic A lobes at
frequencies of tens of MHz will require the low-frequency component of
the Square Kilometer Array (SKA): as noted above, the lowest-frequency
images of \cite{Perley+97} are not good enough for our purposes. It is
therefore important to bear in mind that some structure in the image
can come from differences in the radio and X-ray spectral slope, and
we comment on this in more detail below\footnote{The lack of a strong
  correlation between spectral index and surface brightness in the
  maps of \cite{Perley+97} means that it is difficult to construct a
  model in which {\it all} the differences between the X-ray and radio
  emission come from this difference in the electron energies being
  probed by the two emission mechanisms, as discussed by HC05.}.

\begin{figure}
\includegraphics[width=1.0\linewidth]{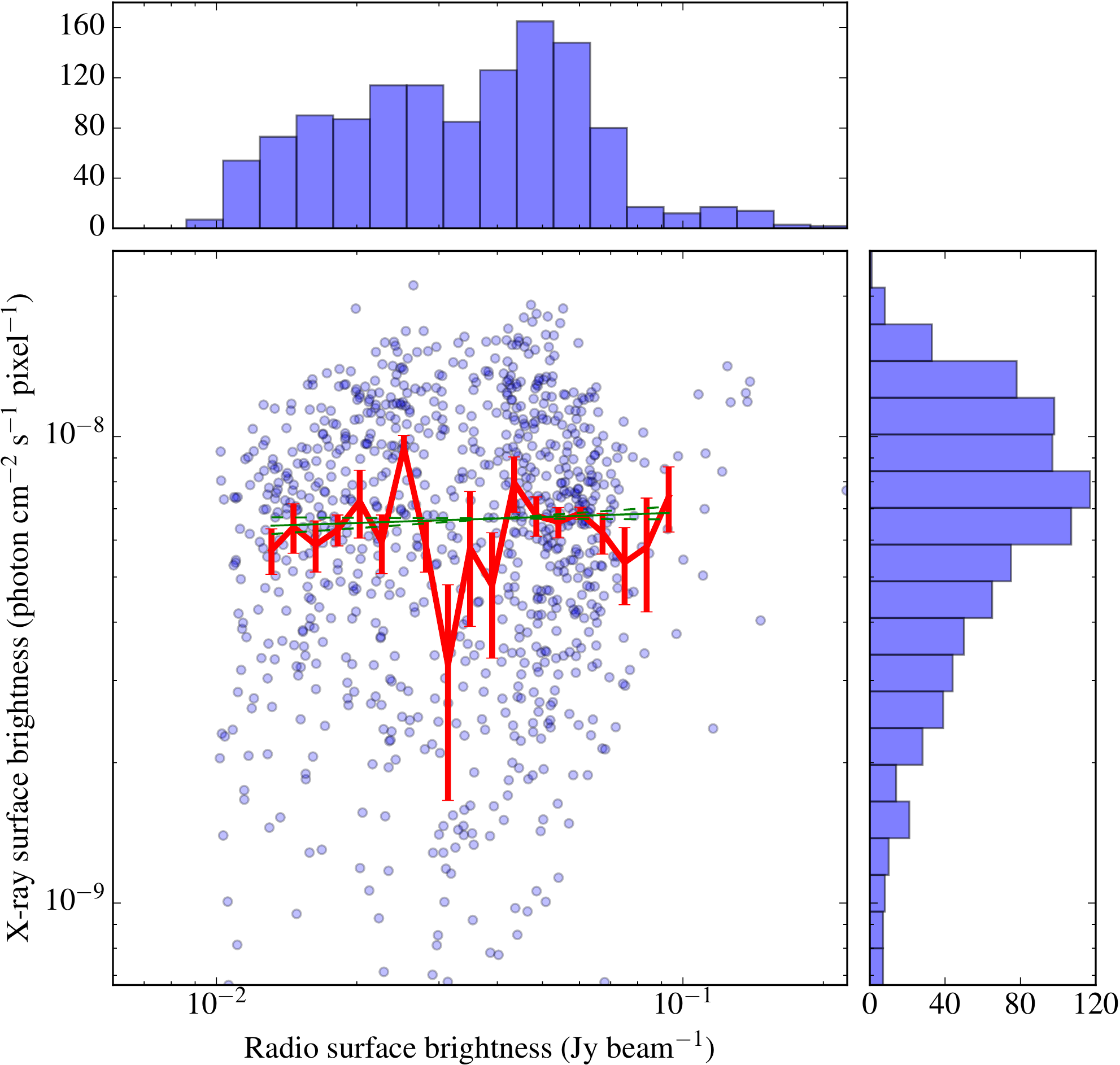}
\caption{Relationship between radio and X-ray surface brightness
  (0.4-2.0 keV) in
  the lobes of Pic A. In the scatter plot at the bottom left, blue points show independent data points as
  discussed in the text; red lines and error bars show an average
  surface brightness in bins of radio flux; the green line shows the
  best-fitting power law describing the median data points in red and
  an estimate of the $1\sigma$ error on its slope. Histograms at the
  top and right show the projections of the radio and X-ray surface
  brightness distributions respectively.}
\label{fig:sb}
\end{figure}

To assess the relationship between X-ray and radio surface brightness
we generated a fluxed X-ray image in the 0.4-2.0 keV band,
exposure-corrected at 1 keV and with 2-arcsec pixels, and convolved it
to a resolution matching that of the radio image. The 1.4-GHz radio
and X-ray images were then regridded to the same resolution. A mask
was applied to the X-ray image to exclude the core, jets, hotspots and
background point sources, and the radio contour at 10 mJy beam$^{-1}$
was used to define the edge of the lobes. Finally, the image was
sampled in distinct regions of $4 \times 4$ pixels ($8 \times 8$
arcsec), taking account of masking, to ensure that each data point was
independent (this is the approximate area of the convolving Gaussian
for both images). Fig.\ \ref{fig:sb} shows the relationship between
radio and X-ray surface brightness derived in this way, and, as can be
seen, there is little or no correlation between them -- the X-ray
surface brightness seems to be independent of the radio, and to be
peaked around a value of $\sim 10^{-8}$ photons cm$^{-2}$ s$^{-1}$
pixel$^{-1}$, though with significant scatter\footnote{Scatter in this
  plot can be the result of statistical noise (Poisson errors) on the
  X-ray emission, or dispersion with a physical origin in the radio
  {\it or} X-ray surface brightness, or both. As it is not easy to
  distinguish between the various sources of dispersion, we do not
  consider the magnitude of dispersion in our analysis.}, over an
order of magnitude variation in radio. This can be seen more clearly
if we take the median X-ray surface brightness in bins of radio
surface brightness, as shown in Fig.\ \ref{fig:sb}. Here we take only
the central 80 per cent of the data in terms of radio surface
brightness to avoid edge effects and structures around the
jet/hotspot, medians are used for robustness to e.g.\ unmasked outlier
points, and errors on the median are derived using bootstrap (which
takes account of both Poisson errors on individual data points before
smoothing and the intrinsic dispersion within a bin). Clearly there is
little correlation even after averaging: the best-fitting power-law
relationship for the binned data points, also plotted in
Fig.\ \ref{fig:sb}, has a power-law index $p = 0.047 \pm 0.038$.
Models in which the electron energy spectrum is constant and the
surface-brightness variation in radio emission are the result of
electron density variations only are conclusively ruled out. We will
discuss the implications of this analysis further in Section
\ref{sec:discussion-lobe}.

\section{Discussion}

\subsection{The jet and counterjet}
\label{sec:disc-jet}

\subsubsection{Spectrum}
\label{sec:disc-jetspectrum}

For the first time we have been able to construct a detailed spectral
profile of the X-ray emission from the jet (Section
\ref{sec:res-jetspectrum}, Fig.\ \ref{fig:profile}). What is remarkable
about this is the uniformity of the spectrum. The photon index does
not deviate significantly from its mean value of $\sim 1.9$ for all of
the 4 arcmin ($\sim 150$ kpc in projection) over which the jet is
observed. This is strong evidence that the dominant emission mechanism
is the same everywhere in the jet. As noted by HC05, the steep X-ray
spectrum is difficult (though not impossible) to reconcile with an
inverse-Compton model for the X-ray emission, since in the beamed
inverse-Compton model the electrons that produce the X-ray emission
have very low energies, much lower even than those producing radio
synchrotron emission, and would be expected to have an energy spectrum
which would be set by the acceleration mechanism: the radio spectral
index of the lobes is flatter than 0.9, as indeed is the 1.4 to
5.5-GHz spectral index of the jets where it can be estimated ($\sim
0.7$) and the radio-optical spectral index (see Gentry \etal\ 2015),
so an inverse-Compton model of the X-ray emission would require the
acceleration process to produce an electron energy spectrum that was
steep at low energies ($\gamma \ll 1000$) and flat at intermediate
energies ($\gamma \sim 10^4$).

\subsubsection{Jet variability}
\label{sec:discussion-jetvariability}

Our programme of jet monitoring has not revealed any new variability
as significant as that reported by M10. It is of course a concern that
the feature most consistent with a short-duration flare appears to
have happened in the very first {\it Chandra} observation, when the
ACIS-S was significantly more sensitive in the soft band than it now
is. (As Fig.\ \ref{fig:jet-lc} shows, the other three, less
significant regions that may be varying in the inner jet are
associated with a broad scatter in the measured flux density as a
function of time rather than a flare at a single epoch.) If we assume
that there are no instrumental explanations for the lack of flares,
then we must conclude that flares at the level reported by M10 are
rare.

Our monitoring campaign does increase the evidence for temporal
variation of the X-ray emission in the jet, given the several
locations of moderately significant variability seen in
Fig.\ \ref{fig:hotspot-variability}. As discussed by M10, the
radiative loss timescales for synchrotron-emitting electrons seem too
long to be relevant to the timescales of the variation that we see
(months to years) although this analysis relies on assumptions about the
magnetic field strength in the jet that cannot be substantiated at
present. If the flaring mechanism is particle acceleration in compact
regions followed by adiabatic expansion, then in principle similar
timescales would be seen whether the emission mechanism is synchrotron
or inverse-Compton emission, though the synchrotron timescales would
be shorter because of the effect of expansion on the magnetic field.

\subsubsection{Jet-counterjet ratio and beaming speed}
\label{sec:ratio}

With a clearly significant detection of the counterjet (Section
\ref{sec:res-jet}), the jet-counterjet ratio is well constrained at
$6.9 \pm 1.2$ (Table \ref{tab:spectra}), consistent with the value
reported by HC05. In a synchrotron model, where the X-ray emission is
isotropic in the rest frame of the jet, and assuming an intrinsically
similar jet and counterjet with the same rest-frame properties, we
expect the jet-counterjet ratio to be given by
\begin{equation}
  {\cal R}  = \left[\frac{1+\beta \cos \theta}{1-\beta \cos
        \theta}\right]^{2+\alpha}
\end{equation}
where $\alpha$ is the spectral index (taken to be 0.9), $\theta$ is
the angle to the line of sight and $\beta = v/c$. We can trivially
solve for the projected speed in the plane of the sky, $\beta
\cos\theta = 0.32$. Since $\beta < 1$ and $\cos \theta < 1$, we
require $\beta \ga 0.3$ and $\theta < 70^\circ$. These are in line
with expectations for a broad-line FRII radio galaxy, which would be
expected from the statistics of unified models to have $\theta <
45^\circ$ and from jet prominence and sidedness statistics to have
apparent $\beta \sim 0.6$ \citep{Wardle+Aaron97,Mullin+Hardcastle09}.
If we assume that $\theta < 45^\circ$, then $\beta \la 0.5$. In the
inverse-Compton model, the speed of the emitting material in the jet is
constrained to be much higher ($\Gamma \ga 5$, i.e. $\beta \approx
1$), the angle to the line of sight must be small (less than a few
degrees) and the emission in the rest frame is anisotropic, increasing
the sidedness asymmetry: even without considering the last factor it
is easy to see that $\cal R$ would be many orders of magnitude higher
than is observed. A one-zone beamed inverse-Compton model is
conclusively ruled out by the counterjet detection alone, unless the
counterjet's emission is produced by some other mechanism (which is
disfavoured by the fact that its X-ray spectrum is identical to that
of the jet).

One other constraint on the jet speed may be provided by the observed
opening angle of $\sim 3^\circ$ (Section \ref{sec:res-jetxray}). For a
freely expanding relativistic jet with no dynamically relevant
magnetic field, we expect $\Gamma\theta\la 1$ where $\Gamma$ is the
jet bulk Lorentz factor, and $\theta$ is the half-opening angle in
radians. The assumption that $\Gamma\theta=1$ has been widely applied
in blazar modelling in the past. However, recent radio studies of
blazar sources indicate that $\Gamma\theta \approx 0.1$--$0.2$
\citep[e.g.,][]{Jorstad+05,Clausen-Brown+13,Zdziarski+15,Saito+15}.
Bearing in mind that the angle to the line of sight of Pic A is
probably $\la 45^\circ$, we have the true half-opening angle $\theta
\la 1^\circ$; this would imply $\Gamma \ga 5$ ($\beta > 0.98$) for the
jet in Pic A. The apparent inconsistency between this bulk speed and
the constraints provided by the jet sidedness analysis above may be
resolved if the jet's dynamics are dominated by a highly relativistic
spine and its emission by a slow-moving boundary layer: we discuss
such a model in more detail in the following subsection.
Alternatively, of course, the jet may not be expanding freely -- see
the discussion of other quasar jets in Section
\ref{sec:res-jet-profile} --, in which case the true value of $\Gamma$
may be lower, or Pic A's jet may differ from those of blazars in some
other way.

It is worth noting that, in contrast to the situation for many of the
core-dominated quasars with bright X-ray jets, we have no {\it direct}
evidence of highly relativistic bulk motions in Pic A on any scale:
the VLBI observations of \cite{Tingay+00} imply at most mildly
superluminal motions. Further VLBI studies of this object would be
valuable.

\subsubsection{Jet physical structure}
\label{sec:disc-jet-origin}

The arguments in Sections \ref{sec:disc-jetspectrum} and \ref{sec:ratio}
strongly disfavour a one-zone inverse-Compton model for the jet X-ray
emission. We regard such a model as untenable and do not
discuss it further. We are therefore forced to the conclusion that
synchrotron emission is responsible for some or all of the X-ray
emission; in individual knots, this conclusion is strongly supported
by the results of Gentry \etal\ (2015), who show that several knots
can be well modelled with a fairly conventional broken-power-law radio
through X-ray synchrotron spectrum.

To understand the origin of the synchrotron X-rays from the jet we
need to consider carefully where the radiating material originates in
FRII jets in general. It is important to note that ruling out the
beamed inverse-Compton model as the origin for the X-ray emission does
not rule out high bulk speeds in the jet, but merely the combination
of high bulk speeds, small angle to the line of sight and plentiful
low-energy electrons in the high-bulk-speed region required for
beamed inverse-Compton to dominate the observed X-rays. This process
may well still operate in other sources, and indeed might dominate the
X-ray emission from Pictor A from a different line of sight. However,
the X-ray jet sidedness seen in Pic A (and the radio jet sidedness and
prominence seen in many other radio galaxies and radio-loud quasars)
would imply that most of the {\it visible} jet emission in the vast
majority of sources comes from moderately relativistic regions of the
jet with $\beta \sim 0.6$ \citep{Mullin+Hardcastle09}. If we
interpreted $\beta \approx 0.6$ as the bulk speed of a homogeneous
jet, then we would have to explain how the jets decelerate to these
speeds, without disruption or extremely obvious dissipation, from the
$\Gamma \ga 10$ implied by VLBI observations of core-dominated
quasars, or even the $\Gamma$ of a few implied by core prominence
variations; we know of no mechanism that can do this. It is much more
natural to consider $\beta \approx 0.6$ (suggestively close to the
sound speed in an ultrarelativistic plasma) as the characteristic
speed of a slow-moving boundary layer which dominates the emission,
while emission from a fast central spine with $\Gamma \ga 10$ (a) is
suppressed by Doppler de-boosting, at least in radio galaxes and
lobe-dominated quasars and (b) may in addition have a substantially
lower rest-frame emissivity. This model (which has been invoked in the
literature previously, see e.g.
\citealt{Bridle+94,Stawarz+Ostrowski02,Hardcastle06,Jester+06,Mullin+Hardcastle09})
has the desirable feature that it does not require the whole jet to
decelerate to moderately relativistic speeds until the jet termination
shock at the hotspot, allowing the interpretation of hotspots as jet
termination shocks even if the jet equation of state is
ultrarelativistic. It also helps to explain the few observations of
edge-brightening in FRII jets \citep[e.g.][]{Swain+98}, and the fact
that FRII jets have polarization-inferred $B$-field pointing along the
jet direction (a natural consequence of shear even if there is a
toroidal field structure in the fast spine). In Pic A, a boundary
layer model is consistent with the low-filling-factor,
non-centrally-brightened appearance of the X-ray jet, and such a model
allows us to reconcile the observed opening angle with the constraints
on its Doppler factor from a sidedness analysis, although we do not
regard this latter point as a particularly compelling argument since
we cannot say whether the jet is really unconfined.

In what follows, we consider various possible origins for the X-rays
based on the boundary-layer model. One consequence of this model is
that the effective jet volume is smaller than its physical volume (as
some of the physical volume is filled with fast-moving,
Doppler-suppressed jet material which we do not (yet) see in any
waveband) and so equipartition magnetic field strength estimates are
lower than they should be; but even the field estimates used by M10
for the whole jet clearly require an in situ acceleration process for
the particles responsible for the synchrotron X-rays. We investigate
the properties required for this acceleration process in the next
subsection.

\subsubsection{Clues to the acceleration process}

We begin by considering the energetics of acceleration. Based on a
very crude broken-power-law model of the radio through X-ray jet
spectrum, the total (observed) radiative jet power is of order $2
\times 10^{35}$ W, or $2 \times 10^{42}$ erg s$^{-1}$ (a number which
is uncertain by a factor of a few in either direction because of the
unknown Doppler factor of the emitting part of the jet within the
constraints imposed by the sidedness analysis of Section
\ref{sec:ratio}). From the results of \cite{Mingo+14} for powerful
3CRR sources, we might expect the jet {\it kinetic} power to be of
order the bolometric AGN power, which is a few $\times 10^{44}$ erg
s$^{-1}$ (Section \ref{sec:AGN}, assuming a bolometric correction from
the X-ray of $\sim 20$). \cite{Mingo+14} use estimates of the jet
power from the method of \cite{Willott+99}, but the correlation they
see between AGN power and jet power has the merit of averaging over
many objects. Alternatively, we can directly use the \cite{Willott+99}
estimate based on the total radio luminosity of Pic A, which gives a
jet power of $\sim 2 \times 10^{45}$ erg s$^{-1}$ if normalized as
described by \cite{Hardcastle+07}, though such estimates depend
strongly on the (unknown) source environment and age
\citep{Hardcastle+Krause13}. Given the rough consistency of these jet
power estimates, we are looking for a mechanism which extracts perhaps
0.1 per cent of the energy transported by the jet and places it in
synchrotron-radiating electrons, predominantly in a slow-moving
boundary layer.

Our observations of Pic A give us some clues about the detailed
particle acceleration mechanism. The flaring behaviour seen by us and
by M10 requires the flaring regions to be small (sub-pc), although the
jet is resolved by {\it Chandra} and therefore several kpc in
diameter. A model in which the observed jet is a boundary layer
alleviates this discrepancy, since the boundary layer may be much
thinner than the observed diameter of the jet, but we still require
either particle acceleration on small scales or rapid adiabatic
compression and expansion of small regions.

Models in which the observed jet emission is a boundary layer make it
somewhat easier to explain the strong variations in jet prominence
both internally to a source (as in Pic A) and within sources, which in
this picture would arise as a result of weaker or stronger
interactions between the invisible `beam' carrying most of the power
and its external environment, causing more or less local acceleration
in the boundary layer. In Pic A, we can speculate that the sudden
reduction in X-ray emissivity about half-way along the jet may be a
result of the jet moving from an environment where it is in contact
with the external thermal environment to one where it is embedded in
the lobes. This would not cause a significant change in the pressure
external to the jet (since the lobes are presumably in approximate
pressure balance with the environment in which they are embedded) but
might well give rise to a change in the rate of dissipation at the
boundary layer. Whether that change would be expected to be
quantitatively anything like what we observe in Pic A is a question
that would require detailed modelling to answer. As we see no direct
evidence for an X-ray-emitting environment in Pic A, there is no
direct observational support for this model other than the sudden
change in jet surface brightness.

We consider some possible physical mechanisms for particle
  acceleration in the following subsections.

\subsubsection{Shocks}

First-order Fermi acceleration at shocks is the first model to
consider, as it is generally thought to be responsible for the
particle acceleration at the hotspots. However, the structure of the
X-ray emission, with elongated regions tens of kpc in length, does not
seem consistent with localized, large-scale shocks such as would be
produced by e.g., reconfinement in the external medium or strong jet
speed variations. In FRI jets, entrainment of dense baryonic material
from the stellar winds of host galaxy stars embedded in the jet is
thought to be responsible for some of the observed kpc-scale
bulk jet deceleration, and \cite{Wykes+15} have recently shown that
diffusive shock acceleration at the many jet/stellar wind boundaries
is energetically capable of producing the observed X-rays in the case
of the well-studied source Cen A, giving {\it distributed},
  shock-related acceleration. But in Pic A and other FRII sources
this cannot be the mechanism responsible for particle acceleration,
since the jet X-ray emission appears on scales where there are
essentially no stars, and certainly too few to intercept the required
fraction of the jet energy. The remaining possibility is some
distributed oblique shocking due to instabilities propagating into the
jet; in principle this could help to explain the quasi-periodic
spacing of the jet knots as well. We cannot rule such a model out out
in the absence of a high-signal-to-noise observation of the jet
boundary, e.g. from still deeper radio observations, but it is not
obvious how it would reproduce the small-scale flaring behaviour of
the jet.

\subsubsection{Shear acceleration}

Shear acceleration \citep{Stawarz+Ostrowski02,Rieger+Duffy04} would in
principle be a natural consequence of the boundary layer model
outlined above. However, as \cite{Stawarz+Ostrowski02} point out, for
electrons of the energies we are discussing here the gyroradius of
electrons in an equipartition field is likely to be many orders of
magnitude smaller than the characteristic size scale of the shear,
greatly reducing the efficiency of the process. M10 estimate an
equipartition field of 1.7 nT for the jet, and though this assumes a
uniformly filled cylindrical geometry, it serves to illustrate the
point: if $\nu$ is the observed synchrotron frequency, then the
electron gyroradius is
  \begin{equation}
    r_g = \sqrt{\frac{\nu}{2\pi}} \frac{m_e^{3/2} c}{q^{3/2} B^{3/2}}
  \end{equation}
where $m_e$ is the electron mass, $c$ the speed of light, and $q$ the
charge on the electron, which for $\nu = 2.4 \times 10^{17}$ Hz gives
$r_g \approx 10^{13}$ m with $B=1.7$ nT, while the jet is transversely
resolved with a radius $R \sim 1$ kpc. Only if the shear layer is very
thin, of order a few times the gyroradius, does shear acceleration
dominate over turbulent acceleration in the jet (cf. eq. 1 of
\citealt{Stawarz+Ostrowski02}). The equipartition field also depends
on the geometry. If the thickness of the boundary layer is $\Delta R$
then, roughly, $B \approx B_{eq} (R/\Delta R)^{2/(p+5)}$ where $p$ is
the power-law electron energy index; thus the gyroradius also
decreases as $\Delta R$ decreases. Setting $p=2.5$ for the sake of the
calculation, we can find a self-consistent value of $\Delta r$ and $B$
which satisfies the condition $r_g/\Delta R \approx 0.3$, but this
still requires $\Delta R/R \approx 10^{-9}$, a completely implausible
geometry. Moreover, the competing process, turbulent acceleration, can
easily be more efficient than assumed by \cite{Stawarz+Ostrowski02};
the proton number density may well be much less than they assume,
giving relativistic Alfv\'en speeds and thus allowing the acceleration
of very high-energy electrons, and requiring $r_g \approx \Delta R$
for shear acceleration to dominate, at which point the assumptions of
the model break down in any case. We discuss turbulent acceleration in
the following subsection.

\subsubsection{Turbulent acceleration and reconnection}

Two remaining widely discussed acceleration mechanisms are stochastic
acceleration in magnetized turbulence \citep[e.g.][and references
  therein]{Stawarz+Petrosian08} and magnetic field line reconnection
\citep[e.g.][]{Sironi+Spitkovsky14,Guo+15}. These possibilities are
not mutually exclusive: the two processes would be expected to operate
together in a turbulent magnetized plasma, and so we consider them
together in this section. Magnetic reconnection is attractive in the
context of Pic A because it naturally produces small-scale, localized
acceleration regions, something which is not necessarily expected in
the case of distributed turbulent acceleration, and these regions are
associated with enhanced magnetic field strengths, increasing their
observability in a synchrotron model; thus it is particularly suitable
for explaining the small-scale flares in the jet. Reconnection can
produce the flat electron energy spectrum that is observed out to the
optical \citep{Gentry+15} without difficulty
\citep{Sironi+Spitkovsky14} and the steeper X-ray spectrum could then
be accounted for in the standard way by losses in a
continuous-injection model (valid so long as the region over which we
integrate observationally exceeds the loss spatial scale, as it does
if the magnetic field strength is close to equipartition). Simulations
such as those of \cite{Sironi+Spitkovsky14} show that the efficiency
of reconnection as an acceleration mechanism depends on the jet
magnetization parameter $\sigma$ (and in particular requires $\sigma >
1$ in the emission regions), and so another attractive feature of the
mechanism is that it can account for variations in the efficiency of
production of X-rays, either within jets as in Pic A's inner and outer
regions or between jets in different sources, by allowing $\sigma$ to
vary. Unfortunately, detailed numerical modelling of reconnection in
the specific physical situation presented by Pic A's jet is
intractable, because of the very large range of spatial scales
involved (see discussion of the electron gyroradius in the
  previous subsection). Relativistic magnetohydrodynamic modelling
(to give the field structure expected at the edge of the jet) together
with some sub-grid model for the microphysics of reconnection might be
able to make predictions about the frequency and intensity of flaring
events that could be compared to our observations.

\subsection{Hotspot spectrum, structure and variability}
\label{sec:discussion-hsvariability}

As we saw in Sections \ref{sec:whotspot} and \ref{sec:ehotspot}, the
hotspot spectra are both reasonably well fitted with power-law spectra
with $\Gamma \approx 1.9$, comparable to the best-fitting power law
for the jet. The X-ray emission in both is much too bright to be
inverse-Compton (specifically synchrotron self-Compton, the favoured
mechanism for hotspots) with an equipartition field strength
\citep{Hardcastle+04}. In the bright W hotspot, there is evidence for
spectral steepening in the X-ray band, with the photon index being 1.9
at the soft end of the band and 2.2 at the high end (the break in
photon index is $0.30 \pm 0.05$): this affects the
compact bright component of the hotspot and may affect the more
diffuse emission to its E as well. For an equipartition field strength
of 16 nT (derived from fits to the data of \cite{Meisenheimer+97},
using a spherical geometry, and thus indicative only) the synchrotron
loss timescale of electrons radiating at an observed frequency
corresponding to 7 keV is $\sim 20$ years, while the physical size of
even the compact component of the hotspot is $\sim 1$ kpc, so losses
of these electrons are inevitable. However, we know from the modelling
of \cite{Meisenheimer+97} that there is a spectral break below the
optical band, at around $10^{13}$ Hz, which should represent the point
at which electron losses start to dominate over transport losses from
the hotspot: electrons radiating at this frequency have a loss
timescale of around $10^4$ years, which is consistent with the
observed physical size of the hotspot if we assume sub-relativistic
outflow from the acceleration region. The further steepening of the
spectrum in the X-ray band must then either represent the first signs
of a high-energy cutoff in the acceleration spectrum, imposed by some
property of the acceleration region, or evidence for additional loss
processes for very high-energy electrons, either synchrotron,
synchrotron self-Compton or adiabiatic. It is not clear whether
additional loss processes could produce as strong a change in the
spectrum as is observed (a steepening in photon index over a factor 10
in photon energy, and so a factor of only 3 in electron energy) but a
high-energy cutoff certainly could do so. Such a model would predict a
steep spectrum in higher-energy observations, a result which can be
tested with forthcoming {\it NuSTAR} observations.

The most unexpected result to emerge from our long-term monitoring is
the apparent decrease in the W hotspot flux by $\sim 10$ per cent over
the period between epochs 7 and 8, on a timescale of only a few months
(Section \ref{sec:whotspot}). Taken at face value, this implies that a
significant fraction of the X-ray emission must come from very small
(sub-pc) regions of the hotspot. \cite{Tingay+08} have shown using
high-resolution radio imaging that a small fraction of the radio flux
in the hotspot (around 2 per cent) is produced by compact structures,
with size scales of tens-hundreds of pc. If the variability we see is
real, though, a larger fraction of the X-ray flux must come from
structures that are even smaller in scale. Our observations provide
some support to the model proposed by \cite{Tingay+08} in which much
of the X-ray emission is produced in the compact components that they
see in the radio, although we note the clear detection of extended
X-ray emission as well, particularly from the `bar' to the E of the
main hotspot structure, which suggests that some of the X-ray emission
is genuinely diffuse. However, the basic picture, in which compact
regions play an important role in high-energy particle acceleration at
the hotspot, clearly explains the observations of Tingay et al. and
the variability we see, and also helps to explain the steepening of
the X-ray spectrum at high energies. The most likely explanation for
the compact regions seen in the radio is that they are due to
localized magnetic field strength overdensities within the hotspot:
such structures will be privileged sites for particle acceleration but
will also necessarily be transient, since their excess of magnetic
field strength will drive expansion (as pointed out by Tingay \etal).
Variability in the X-ray and losses in the X-ray regime over and above
those predicted from a simple one-zone model are both naturally
expected in this picture. We might also expect some variability in the
radio, and it would be interesting to investigate both the integrated
variability of the hotspot and any variability in the small-scale
radio features seen by Tingay \etal

The W hotspot, and quite possibly also the E hotspot if we accept the
possible association between components R1 and X2, show offsets on kpc
scales between the X-ray and radio peaks. These offsets have been seen
in many other sources (see
\cite[e.g.][]{Hardcastle+02,Hardcastle+07,Perlman+10,Orienti+12}), and
are always in the sense that the X-ray emission is further upstream
(closer to the nucleus) than the radio.
\cite{Georganopoulos+Kazanas03} proposed that some of the X-ray
emission from hotspots might be produced by inverse-Compton
upscattering by jet material of synchrotron photons from the shocked
region: this predicts an offset in the sense (though not necessarily
of the magnitude) that is observed. However, as noted by
\cite{Hardcastle+07}, such a model fails to explain the offsets in
sources aligned close to the plane of the sky, or in double hotpots.
In Pic A, the lack of any spectral difference between the bright peak
emission, where the offset is seen, and the more diffuse bar emission
(Section \ref{sec:whotspot}) also argues against a role for
inverse-Compton emission, which would be expected to have a flat
spectrum. Morphologically, the bar region as seen in the X-ray has
structure very similar to what is seen in the optical, where
polarization clearly implies a synchrotron origin for the emission
\citep[e.g.][]{Wagner+01,Saxton+02}; both the optical and X-ray data,
interpreted as synchrotron, require distributed, in situ particle
acceleration in the W hotspot.

Finally, we draw attention to the very different structural properties
of the E hotspot (Section \ref{sec:ehotspot}), which contains mostly
diffuse X-ray emission, extended over tens of kpc with only a few,
faint (and possibly unrelated) compact components. Pic A is one of a
number of well-studied broad-line FRIIs to show this difference
between the jet-side and counterjet-side hotspots (see
\cite{Hardcastle+07} for the examples of 3C\,227 and 3C\,390.3). The
natural explanation for this (and for various radio properties of
large samples of hotspots, e.g. \citealt{Bridle+94}) is that there is
some relativistic bulk motion downstream of the jet termination shock
\citep{Laing89,Komissarov+Falle96}, an idea that has been discussed
previously in the context of Pic A (HC05; \citealt{Tingay+08}), and
that this suppresses on the counterjet side the bright compact
emission associated with the jet termination itself, where the
direction of the flow is still away from the observer, but enhances
diffuse emission, which occurs in the backflow, directed towards the
observer. Some consequences of this model were discussed by
\cite{Bridle+94}. Here we simply note that this is much easier to
arrange (and likely to be much more significant for the observed
properties of hotspots) if the mean bulk jet speed significantly
exceeds the `beaming speed' of $\sim 0.6c$, as in the models discussed
in the previous subsection. In such a model the {\it hotspot}
sidedness ratio (considering the compact component directly downstream
from the termination shock) can be comparable to or even exceed the
{\it jet} sidedness ratio, which arises from relatively slow-moving
material in a boundary layer: this must be the case in Pic A if the
terminations of the two jets are at all similar in the rest frame.

\subsection{Magnetic field and electron distribution in the lobes}
\label{sec:discussion-lobe}

Pic A's lobes, as the brightest clearly detected inverse-Compton lobes
in the sky (Section \ref{sec:intro-lobes}), represent an excellent
laboratory for studies of the nature of the lobe plasma. In Section
\ref{sec:lobes} we measured a flux density for the lobes compatible
with the earlier estimates of HC05 and a flat low-energy photon index
of $1.57 \pm 0.04$ (statistical errors only), which is consistent with
expectations for particle acceleration at strong shocks if the lobe
emission is inverse-Compton from scattering of the CMB. Using simple
one-zone lobe models and the code of \cite{Hardcastle+98}, the flux we
measure implies a mean magnetic field strength of around 0.4 nT in the
lobes, a factor $\sim 1.5$ below the equipartition field strength. Pic
A is very similar to other FRIIs in showing this slight departure from
equipartition \citep{Croston+05}. The pressure from the radiating
components of the lobes (electrons and field) would then be around
$10^{-13}$ Pa, which would provide pressure balance with an external
thermal atmosphere only if the environment is very poor.

In Section \ref{sec:lobes} we drew attention to the very poor
correlation between the X-ray and radio surface brightness in the
lobes, which is clearly inconsistent with a model in which the
variation in synchrotron surface brightness is caused by electron
density variations in a constant magnetic field. Here we
investigate\footnote{An earlier version of the modelling process
  described in this section was discussed by \cite{Goodger10}, with
  similar conclusions.} the constraints placed by our observations on
the family of models with the relationship between the local electron
energy spectrum normalization and magnetic field described by a
parameter $s$ \citep{Eilek89},
\begin{equation}
N_0 \propto \left(\frac{B}{B_0}\right)^s
\label{s-definition}
\end{equation}
Here $s=0$ corresponds to the uniform electron density case discussed
above, $s=2$ corresponds to local equipartition, and we can
conveniently denote the case with uniform field and arbitrarily
varying electron density, which we have already ruled out, as $s =
\infty$.

\begin{table}
\caption{Median and (10th, 90th) percentile radio/X-ray correlation
  slopes $p$ for simulated lobes as a function of the field/electron
  correlation parameter $s$ described in the text}
\label{table:sims}
\begin{tabular}{lrrr}
\hline
$s$&\multicolumn{3}{c}{Power-law slope $p$}\\
&Median&10th percentile&90th percentile\\
\hline
0.00 & 0.169 & 0.009 & 0.232 \\
0.25 & 0.183 & 0.081 & 0.247 \\
0.50 & 0.229 & 0.136 & 0.272 \\
0.75 & 0.216 & 0.143 & 0.278 \\
1.00 & 0.233 & 0.153 & 0.334 \\
1.50 & 0.259 & 0.192 & 0.350 \\
2.00 & 0.308 & 0.191 & 0.402 \\
\hline
\end{tabular}
\end{table}

To do this we carry out a number of realizations of a spherical lobe
with a Gaussian random magnetic field having a Kolmogorov power
spectrum, using the code described by \cite{Hardcastle13} -- we
verified that this power spectrum for the field leads to a power
spectrum in projected synchrotron emission that is consistent with
what is observed in Pic A. The electron energy spectrum is assumed to
be the same throughout, and, importantly, is chosen to reproduce the
integrated spectrum of the lobes of Pic A (so that there is some
spectral steepening in the radio with respect to the low-energy
electrons that produce the inverse-Compton emission). This is
necessary because the effect on emissivity of varying the magnetic
field at a given observing frequency depends strongly on the local
spectral index. The electron spectrum normalization is taken to depend
everywhere on the local value of $B$ as described by
eq.\ \ref{s-definition}. Synchrotron and inverse-Compton visualization
were then carried out as described by \cite{Hardcastle13}, the images
were resampled to give the same number of independent data points as
in the real images, and the slope of the power-law radio/X-ray surface
brightness correlation was determined by binning in radio surface
brightness and finding the errors via bootstrap in exactly the manner
carried out for the real data in Section \ref{sec:lobes}. This process
was repeated for a number of different discrete values of $s$, and,
for each $s$, repeated many times in order to form some kind of
average over the randomly generated visualizations.

Results are given in Table \ref{table:sims}, where we show both the
median correlation slopes and the 10th and 90th percentile values to
give some indication of the breadth of the distribution. We see, as
expected, that higher values of $s$ give rise to stronger
correlations. The median correlation is always positive -- simple
geometrical effects guarantee that there will always be some positive
correlation -- but straight away we can see that only low values of
$s$ produce power-law slopes as flat as the one actually observed
($p=0.047 \pm 0.038$) In fact, only $s=0$ produces a slope as flat as
the one observed in more than 10 per cent of the simulations (Table
\ref{table:sims}). This conclusion is robust to the introduction of
uncorrelated noise into the electron densities, because this averages
out both when integrating along a line of sight and when binning in
radio surface brightness: for example, if we (unrealistically) add
Gaussian noise with $\sigma = 0.5$ times the mean electron density to
each volume element, truncating the electron density at zero when
necessary, the results of Table \ref{table:sims} are essentially
unaltered, although the scatter on a plot such as that of
Fig.\ \ref{fig:sb} is obviously increased. Even taking account of the
errors on the fitted value of $p$, it is very hard to see how values
of $s\ge 1$ can be reconciled with the data, and we suggest that this
is additional strong evidence that the filamentary structures in lobes
in general, and Pic A in particular, are dominated by variations in
magnetic field strength with little correlated variation of the
electron density. This conclusion is consistent with expectations from
numerical modelling \citep{Hardcastle+Krause14}.

\section{Summary and conclusions}

We have presented results from a long-term, sensitive programme of
{\it Chandra} observations of the broad-line radio galaxy Pictor A,
which give us unparallelled sensitivity to variability in compact
components of the jet and hotspot coupled with by far the deepest view
of inverse-Compton emission from a radio galaxy's lobes. Key results
may be summarized as follows:

\begin{enumerate}
\item Both a jet and counterjet are detected extending all the way
  from the nucleus to the hotspot region (Section \ref{sec:res-jet}). The jet/counterjet flux
  ratio is completely incompatible with a beamed inverse-Compton
  origin for the X-rays, assuming the two jets are intrinsically
  identical: together with the now extremely well-constrained steep
  spectrum of the jet and with arguments from the detailed broad-band jet spectrum
  \citep{Gentry+15} we conclude that the jets in Pic A are clearly
  synchrotron in origin (Section \ref{sec:disc-jet}).

\item We have not seen any further flares at the level reported by
  M10, but there is further evidence for low-level, short-term
  variability in the jet (Section \ref{sec:res-jetvar}). At the same
  time, there is no evidence for any change in jet spectral index as a
  function of position, strongly arguing that the acceleration
  mechanism is constant along the jet, although acceleration {\it
    efficiency} may well vary (Section \ref{sec:res-jetspectrum}). We
  suggest that distributed, localized particle acceleration due to
  magnetic field reconnection may provide the best explanation for the
  observations (Section \ref{sec:disc-jet-origin}).

\item The well-studied bright W X-ray hotspot is shown to have a
  spectral steepening across the band, arguing for a spectral cutoff
  or at least significant losses in the high-energy electrons, and to
  be significantly offset with respect to the radio, as seen in a
  number of other hotspots. More importantly, we have found the first
  evidence for hotspot temporal variability on timescales of months to
  years (Section \ref{sec:whotspot}). These timescales correspond to
  spatial scales much smaller than the physical size of the hotspot,
  and we argue (Section \ref{sec:discussion-hsvariability}) that this
  implies a significant contribution to the hotspot X-ray flux from
  one or a few very compact, bright regions, perhaps related to the
  compact radio sources seen by \cite{Tingay+08}. We suggest that
  these are transient features caused by very high localized magnetic
  field energy density, presumably a result of shock compression of
  the already complex magnetic field structure that is transported up the
  jet: if so, they would be expected to be variable at some level at
  all wavebands and it would be very interesting to monitor the
  hotspot flux evolution in the radio.

\item In the bright inverse-Compton lobes we show that there is a very
  poor correlation between radio and X-ray surface brightness (Section
  \ref{sec:lobes}, which
  is consistent with models in which the electron density is
  relatively uniform and the variations in radio surface brightness
  are largely due to spatial magnetic field variations (Section
  \ref{sec:discussion-lobe}). 
  
\end{enumerate}

What implications do these results have for other radio-loud AGN? As
noted above (Section \ref{sec:intro}) Pic A's X-ray jet is exceptional
among jets identified as having a clear synchrotron origin in that it
is detected for the whole of its length rather than as a few isolated
`jet knots'. This, however, seems likely to be at least partly the
result of modest Doppler boosting and of the proximity of Pic A rather
than because it is physically unusual in some way. Another broad-line
radio galaxy at a similar though slightly larger distance, 3C\,111,
shows a similar X-ray jet (Perlman \etal\ in preparation). The mechanisms that
we have discussed for particle acceleration in the jet should be
capable of operating in all FRII jets, though perhaps (particularly if
our discussion of reconnection-related acceleration above is correct)
with a wide range of intrinsic efficiencies. Therefore we would expect
all FRII synchrotron jets to show similar behaviour: for example, we
would expect steep ($\sim 2$) photon indices in the X-ray with little
variation along the jet. Unfortunately there are few sources with
which to test this prediction, but, for example, it is consistent with
observations of the knots of 3C\,353 \citep{Kataoka+08}. A
reconnection model offers a natural explanation for the tendency of
X-ray jets to be brighter relative to the radio closer to the AGN,
seen in Pic A, in 3C\,353 and in many quasar jets
\citep{Marshall+01,Sambruna+04,Hardcastle06}, if we assume that the
boundary layer of the jet is initially more strongly magnetized and/or
has more field reversals, and that these parameters are affected by
dissipation along the jet. Models of this kind, with {\it in situ}
acceleration of particles in many small regions with varying
efficiency, do not have to produce a smooth broad-band
synchrotron spectrum with a monotonically increasing spectral index as
a function of frequency when integrated over regions much larger than
the loss spatial scale (though they may do so, as we see in the jet of
Cen A and in the individual knots in Pic A detected in the optical by
\citealt{Gentry+15}). Consequently, synchrotron emission cannot be ruled out
if, for example, a spectral flattening (${\rm d}\alpha/{\rm d}\nu
  > 0$) is observed at some point.

It is important to emphasise, however, that nothing in our results
rules out the alternative boosted inverse-Compton model for some or
all of the X-ray emission in other beamed systems. In the picture we
have outlined above, the X-ray emission will in fact always be a
combination of the two processes. What we see will depend on both the
intrinsic emissivity of the fast and slow components of the jets --
which in the synchrotron case, we suggest, is dependent on local
conditions in the boundary layer, and in the inverse-Compton case by
the effectively unknown properties of the low-energy electrons in the
fast spine of the jet -- and on beaming, which will Doppler-suppress
emission from the fast spine in all but the most closely aligned jets.
High-resolution radio observations of large samples of FRII jets are
required to test the model in which these systems have velocity
structure and to investigate the properties of the fast-moving
component if it exists. These may be provided by Jansky VLA and
e-MERLIN observations in the coming years.

\section*{Acknowledgments}

We thank Rick Perley for providing radio images from \cite{Perley+97},
Dan Schwartz and Dan Harris for helpful comments on an earlier draft
of the paper, and the referee, Robert Laing, for numerous
  valuable comments on the content and presentation. This research
has made use of data obtained from the {\it Chandra} Data Archive, and
software provided by the {\it Chandra} X-ray Center (CXC) in the
application packages CIAO and Sherpa. The Australia Telescope Compact
Array is part of the Australia Telescope National Facility which is
funded by the Commonwealth of Australia for operation as a National
Facility managed by CSIRO. The National Radio Astronomy Observatory is
a facility of the National Science Foundation operated under
cooperative agreement by Associated Universities, Inc. This research
made use of Astropy, a community-developed core Python package for
astronomy \citep{AstropyCollaboration13} hosted at
http://www.astropy.org , and of APLpy, an open-source plotting package
for Python hosted at http://aplpy.github.com . The Centre for All-sky
Astrophysics (CAASTRO) is an Australian Research Council Centre of
Excellence, funded by grant CE110001020. MJH, JLG and JHC acknowledge
support from the UK STFC via grants ST/M001008/1 and ST/M001326/1.
\L.S. was supported by Polish NSC grant DEC-2012/04/A/ST9/00083.

\bibliographystyle{mnras}
\renewcommand{\refname}{REFERENCES}
\bibliography{../bib/mjh,../bib/cards}

\end{document}